\documentclass{aastex62}

\received{}
\revised{}
\accepted{}
\submitjournal{}

\shorttitle{Sample article}
\shortauthors{Salehi et al.}
\begin{document}

\title{The cosmological bulk flow in QCDM model: (In)consistency with $\Lambda CDM$}

\correspondingauthor{Amin Salehi}
\email{salehi.a@lu.ac.ir}

\author[0000-0002-0786-7307]{A. Salehi, M. Yarahmadi and S. Fathi}
\affil{Department of Physics, Lorestan University , Lorestan, Iran \\
}



\begin{abstract}
We study the bulk flow of the local universe using Type Ia supernova data by considering a class of cosmological model which is spatially flat,(FRW) space-times and contains cold dark matter and $Q$ component (QCDM
models) of the fluid as a scalar field, with self interactions determined by a potential $V(Q)=V_{0}Exp(-\lambda Q)$ evolving in Universe. We use different cumulative redshift slices of the  Union 2 catalogue. A maximum-likelihood analysis of peculiar velocities confirms that at low redshift $0.015 <z<0.1$, bulk flow is moving towards the $(l; b) = (302^{o}\pm20^{o}; 3^{o}\pm10^{o})$ direction with $v _{bulk} = 240\pm 25kms^{-1} $ velocity. This direction is aligned  with direction of (SSC) and agreement with a number previous studies at $(1-\sigma)$, however for high redshift $0.1 <z<0.2$, we get $v _{bulk} = 1000\pm 25kms^{-1} $ towards the $(l; b) = (254^{+16^{o}}_{-14^{o}}; 6^{+7^{o}}_{-10^{o}})$. This indicates that for low redshift our results are approximately consistent with the $\Lambda CDM$ model with the latest WMAP best fit cosmological parameters however for high redshift they are in disagreement of $\Lambda CDM$ and support the results of previous studies such as Kashlinsky et. al, which report the large bulk flow for the Universe. We can conclude that, in $QCDM$ model, at small scales, fluctuations of the dark energy are damped and do not enter
in the evolution equation for the perturbations in the pressureless matter, while at very large scales $(\sim > 100 h^{-1}Mpc)$, they leaving an imprint on the microwave background anisotropy.

\end{abstract}



\section{Introduction} \label{sec:intro}
The Dipole Anisotropy (DA) is the best interpreted as motion of our Local Group (LG) with amplitude of $627\pm22 kms^{-1}$, with respect to the Cosmic Microwave Background (CMB)  towards preferred direction
  $(l, b) = (276^{o}\pm3^{o}\, 30^{o}\pm3^{o})$ in galactic coordinates \citep{Kogut}.
The measurements of the dipole anisotropy of the cosmic
microwave background (CMB) have a long
history \citep{Lineweaver}. The first measurement
 was made by \citep{Conklin} using a ground-based differential radiometer working at 8 GHz and confirmed by the results of \citep{Henry}.
These studies followed by several studies \citep{Lineweaver} and consequently more precise determination was provided by \citep{Smoot}. It was suspected that the gravitational attraction towards a nearby
overdensity might be responsible for the LG motion. In this respect, the studies was focused on investigation the dipole induced by the gravitational influence of structures
in our Local Universe and comparison it with CMB dipole. The first attempt was made
by \citep{Yahil} using the Revised Shapley–Ames (RSA) catalog
. This
effort was traced by several studies \citep{Davis0}; \citep{Davis}; \citep{Shaya}; \citep{Yahil2}; \citep{Aaronson};
\citep{Villumsen}; \citep{Dressier}; \citep{Lynden}; \citep{Robinson}; \citep{Lahav}; \citep{Strauss}; \citep{Hudson}. At first, it was thought that the virgo cluster might be the source of this motion, however the
direct measurements of the Virgocentric flow showed that this motion is not directly pointed at Virgo, and further regions of over density are required to fully explain the DA \citep{Davis} and \citep{Villumsen}. As pointed out by \citep{Shaya}; \citep{Tammann}; \citep{Aaronson}, the vector difference between Virgocentric flow and the DA points in the general direction of the Hydra- Centaurus region. This indicates that the Local Group is feeling the attraction of its nearest mass concentration, the Hydra - Centaurus supercluster. Additional analysis by \citep{Lahav}showed that the general mass distribution within a radius of $4000kms^{-1}$ might be responsible for the acceleration of the Local Group. Fuller sky coverage later revealed disconcertingly large positive velocity residuals (motion away from the observer) in the Hydra - Centaurus region [e.g. \citep{Dressier}]. If Hydra - Centaurus is moving with respect to the CMB, then it cannot be the sole source of the observed DA, and a more distant mass concentration is required if the motion is gravitational in origin. More analysis by, \citep{Lynden-Bell} led to a model in which bulk flow was replaced by flows that are driven by a rather large mass concentration (the " Great Attractor ") which lies beyond Hydra - Centaurus at a kinematic distance of $4350kms^{-1}$. Thus, the Local Group feels the accelerations of both the Virgo Cluster and the GA. A number of authors claimed that this motion is not due to nearby sources, such as the Great Attractor (at a distance of $40 h^{-1} Mpc$),
but rather to sources at greater depths that have yet to be fully identified . For example,\citep{Kocevski} found that the GA only accounts for 44\% of the dipole anisotropy in a large X-ray cluster sample, with the rest evidently caused by more distant sources such as the Shapley Supercluster (SSC) at a distance of $105-165h^{-1}Mpc$ $(0.035 < z < 0.055)$ in the direction $ (l, b) = (306.44^{o}, 29.71^{o})$ . A Large number of studies confirm that one might need to go well beyond $150 h^{-1} Mpc$ in order to fully recover the dipole motion \citep{Lavaux}; \citep{Shapley}; \citep{Scaramella}; \citep{Raychaudhury}.\

Over larger distances,\citep{Kashlinsky} reported a coherent bulk flow out to $d\geq 300 h^{-1}Mpc$ by analyzing the X-ray galaxy clusters using kinematic Sunyaev - Zeldovich (kSZ) effect. In their latest results \citep{Kashlinsky}-\citep{Kashlinsky4}, the bulk flow was found pointing to $(l, b) = (283^{o}\pm14^{o}, 12^{o}\pm14^{o})$ with the peculiar velocity up to $~ 1000 km s^{-1}$ at the scales up to $\sim800 h^{-1}Mpc$. A bulk flows  with this amplitude on such a large scale can not predicted by $\Lambda CDM$ cosmology. In this case, it seems impossible to generate cosmologically consistent results simply by tinkering with the parameters of$\Lambda CDM$; instead a wholesale revision of the model would be called for \citep{Watkins}.
Here we are going to investigate the bulk flow in a model in which the more usual cosmological constant is replaced with a dynamical, time dependent
component which contain cold dark matter and "quintessence"or  the Q-component (QCDM).The
basic idea of the quintessence model bases up on a scalar field $Q$ that slowly evolves down its
potential $V(Q)$.
 While $QCDM$ and $\Lambda CDM$ both  provide a good fit to the
observation data. However, $\Lambda CDM$  model suffer from several problems \citep{Weinberg}-\citep{Peebles}, also $QCDM$ has advantages in fitting constraints from
high red shift supernovae, gravitational lensing, and
structure formation at large red shift $(z \sim 5)$ and at very large scales $ (\sim > 100 h^{-1}Mpc)$, Quintessence clusters gravitationally, leaving an imprint on the microwave background anisotropy \citep{Caldwell}.The spatial inhomogeneities in $Q$ evolve over time due to the gravitational interaction between $Q$ and clustering matter (Caldwell et al. 1998). The perturbations are important because they can leave a distinguishable imprint on the CMB and large-scale structure.
We consider a scalar field with an exponential potential energy density $V(Q)=V_{0}exp(-\lambda\kappa Q)$ evolving in a
spatially -flat(FRW) universe containing a fluid with barotropic equation of state $P_{\gamma}=(\gamma-1)\rho_{\gamma}$, where $\gamma$ is a constant, $0\leq\gamma\leq2$, $ \kappa^{2}\equiv 8\pi G$ and $\lambda$ is a constant . The total energy density and presume of a homogeneous scalar field are
\begin{equation}\label{is}
\rho_{Q}=\frac{1}{2}\dot{Q}+V(Q),\ \ \
P_{Q}=\frac{1}{2}\dot{Q}-V(Q)
\end{equation}
The other intriguing feature of slowly rolling quintessence$(\frac{1}{2}\dot{Q}\ll V(Q)$ is that it behaves like variable cosmological “constant”\citep{Ratra}, slowly evolve with time and to the lowest order approximation, dark energy behaves like a cosmological constant(Its EOS,$w=\frac{P_{Q}}{\rho_{Q}}\approx-1$).
The Klein-Gordon equation of the quintessence field is

\begin{equation}
\ddot{Q}+3H\dot{Q}+\frac{d V}{d Q}=0\label{is3}\\
\end{equation}
The evolution equations for the model are
\begin{equation}
H^{2}=\frac{\kappa^{2}}{3}(\rho_{\gamma}+\frac{1}{2}\dot{Q}^{2}+V)\label{is1}
\end{equation}
\begin{equation}
\dot{H}=-\frac{\kappa^{2}}{2}(\rho_{\gamma}+p_{\gamma}+\dot{Q}^{2})\label{is2}
\end{equation}
\begin{equation}
\dot{\rho}_{\gamma}=-3H(\rho_{\gamma}+p_{\gamma})\label{is4}
\end{equation}

\section{Theoretical Calculation of Bulk flow} \label{sec:style}
For the study of anisotropies and bulk flows present in SN Ia data the dipole fit (DF) method based on, \citep{Bonvin}is used to determine the bulk flow velocity in redshift shells
\begin{equation}\label{is}
d_{L}(z,\upsilon_{bulk},\theta)=d^{0}_{L}(z)+d^{dipole}_{L}(z,\upsilon_{bulk},\theta),
\end{equation}
where
\begin{equation}\label{is00}
d^{0}_{L}(z)=c(1+z)\int^{z}_{0}\frac{dz^{\prime}}{H(z^{\prime})},
\end{equation}
 $z$, is the cosmological redshift, $\upsilon_{DF}$ is the dipole velocity range ``$\theta$`` is the angle between the sight line and H(z) represents the Hubble parameter.
The dipole term $d^{dipole}_{L}(z,\upsilon_{DF},\theta)$ can be written as
\begin{equation}\label{isb}
d^{(dipole)}_{L}(z,\upsilon_{bulk},\theta)=\frac{\upsilon_{DF}(1+z)^{2}}{H(z)}.\cos(\theta).
\end{equation}

Several authors have attempted to derive an expression for luminosity distance in a perturbed RW Universe .\citep{Sasaki} has studied the luminosity distance as function of redshift for a general perturbed space-time. Sasaki’s analysis gave an explicit expression for an Einstein de-Sitter universe.
An explicit expression for the luminosity distance was derived by \citep{Pyne} and was later corrected by \citep{Hui}. In this study, they, have derived an expression for the luminosity distance fluctuation that is accurate to first order, and has a number of terms which can be loosely divided into four categories: peculiar motion (first line), gravitational lensing (second line), gravitational redshift (third line) and integrated Sachs - Wolfe (fourth and fifth lines) (see Eq. (C21)) of \citep{Hui}. They have shown that among all first order terms, the peculiar motion and lensing terms dominate in realistic applications[ see (Eq.18) of \citep{Hui}].
This is because we are generally interested in fluctuations on scales smaller than the horizon. The high redshift SN surveys generally cover a small fraction of the sky while the low redshift surveys, even though they cover a significant fraction of the sky, do not extend out to a sufficient depth to be sensitive to horizon scale fluctuations. Further discussions can be found in [Appendex C of \citep{Hui}].\\
The above studies provided a unified treatment valid at both low and high redshift and revealed clearly how the lensing and peculiar velocity effects come to dominate at high $(z>~0.1)$ and low redshifts $(z<~0.1)$ respectively. Following this studies,\citep{Bolejko}have noted that the standard lensing convergence effect is overwhelmed at low redshifts by a relativistic Doppler term that is typically neglected.\\
By introducing the following dimensionless variables
\begin{equation}\label{is}
\Theta_{1}\equiv\frac{\kappa\dot{Q}}{\sqrt{6}H} , \Theta_{2}\equiv\frac{\kappa\sqrt{V}}{\sqrt{3}H},\Theta_{3}\equiv\frac{\kappa\sqrt{\rho_{\gamma}}}{\sqrt{3}H}
\end{equation}
It is possible to write the evolution equations as a phase plane autonomous system as

\begin{equation}
\frac{d\Theta_{1}}{dN}=-3\Theta_{1}+\sqrt{\frac{3}{2}}\lambda \Theta_{2}^{2}+\frac{3}{2}\Theta_{1}[2\Theta_{1}^{2}+\gamma(1-\Theta_{1}^{2}-\Theta_{2}^{2})]\label{f1}\\
\end{equation}
\begin{equation}
\frac{d\Theta_{2}}{dN}=-\lambda\sqrt{\frac{3}{2}} \Theta_{1}\Theta_{2}+\frac{3}{2}\Theta_{2}  [2\Theta_{1}^{2}+\gamma(1-\Theta_{1}^{2}
-\Theta_{2}^{2})]\label{f2}
\end{equation}
Where, $N=lna$. Also the important parameter, $\frac{\dot{H}}{H^{2}}$, in terms of new variables will be
\begin{equation}\label{hd}
\frac{\dot{H}}{H^{2}}=-\frac{3}{2}\Big(2\Theta_{1}^{2}+\gamma(1-\Theta_{1}^{2}-\Theta_{2}^{2})\Big)
\end{equation}
The above parameter is one of the useful parameters which can relate the theoretical model with observation. In fact by using this parameters and introducing two new  variables
$\Gamma=H$ and $\vartheta=d_{L}^{0}(z)$, we can convert the equation (\ref{is00}) to the two equivalent differential equations as follows

\begin{equation}
\frac{d\vartheta}{dN}=-\Big(\vartheta+\frac{e^{2N}}{\Gamma}\Big)\label{d1}\\
\end{equation}
\begin{equation}
\frac{d\Gamma}{dN}=\varepsilon\Gamma\label{h1}
\end{equation}
Where, we have supposed, $\varepsilon=\frac{\dot{H}}{H^{2}}$. Thus in order to find the bulk flow velocity we need to solve the set of equations (\ref{d1}, \ref{h1}) and (\ref{f1}, \ref{f2})
simultaneously as a equations set as follows
\begin{equation}
\frac{d\Theta_{1}}{dN}=-3\Theta_{1}+\sqrt{\frac{3}{2}}\lambda
\Theta_{2}^{2}+\frac{3}{2}\Theta_{1}[2\Theta_{1}^{2}+\gamma(1-\Theta_{1}^{2}-\Theta_{2}^{2})]\label{f11}\\
\end{equation}
\begin{equation}
\frac{d\Theta_{2}}{dN}=-\lambda\sqrt{\frac{3}{2}} \Theta_{1}\Theta_{2}+\frac{3}{2}\Theta_{2}  [2\Theta_{1}^{2}+\gamma(1-\Theta_{1}^{2}
-\Theta_{2}^{2})]\label{f22}\\
\end{equation}
\begin{equation}
\frac{d\vartheta}{dN}=-\Big(\vartheta+\frac{e^{2N}}{\Gamma}\Big)\label{d11}\\
\end{equation}
\begin{equation}
\frac{d\Gamma}{dN}=\varepsilon\Gamma\label{h11}
\end{equation}

\section{Numerical analysis}
In this paper, we use the Union2 compilation \citep{Amanullah} of 577 SNe  and covers the redshift range $0.015 < z <1.4$  \\

In order to fit the Union2 dataset to a dipole
anisotropy we proceed as follows\\
$\bullet$ We convert the equatorial coordinates of each
supernovae to galactic coordinates.\\
$\bullet$ We find the Cartesian coordinates of the unit vectors $\hat{n}_{i}$ corresponding to each supernovae with galactic

\begin{equation}\label{is}
\hat{n}_{i}=cos(l_{i})sin(b_{i})\hat{i}+sin(l_{i})sin(b_{i})\hat{j}+cos(b_{i})\hat{k}
\end{equation}
where $(l_{i},b_{i})$  is the galactic coordinates of the (i)th supernova . Also $\hat{p}$ is the unit vector in direction of dipole then:
\begin{equation}\label{is}
\hat{p}=cos(l)sin(b)+sin(l)sin(b)\hat{j}+cos(b)\hat{k}
\end{equation}
which $(l,b)$ is  bulk flow direction in galactic coordinate , so
\begin{equation}\label{is}
cos\theta_{i}=(\hat{n}_{i}.\hat{p})=cos(l)sin(b)cos(l_{i})sin(b_{i})+sin(l)sin(b)sin(l_{i})sin(b_{i})+cos(b)cos(b_{i})
\end{equation}
We can constrain on the direction and bulk flow velocity by minimizing the $\chi^{2}$, which is
constructed as follow.
\begin{equation}\label{is}
\chi^{2}=\sum_{i}\frac{|\mu_{i}-5\log_{10}((d^{0}_{L}(z_{i})-d^{dipole}_{L}(z,\upsilon_{DF},\theta_{i})/10 pc|^{2}}{\sigma^{2}_{i}}
\end{equation}
Where,
\begin{eqnarray}\label{distancem}
\mu _{i}=5\log_{10} d_{L}(z)+42.384-5\log_{10} h_{0}
\end{eqnarray}
 The numerical analysis for different redshift ranges is as follows:

\subsection{Numerical analysis for redshift $0.015<z<0.035$}
\begin{figure*}
\gridline{\fig{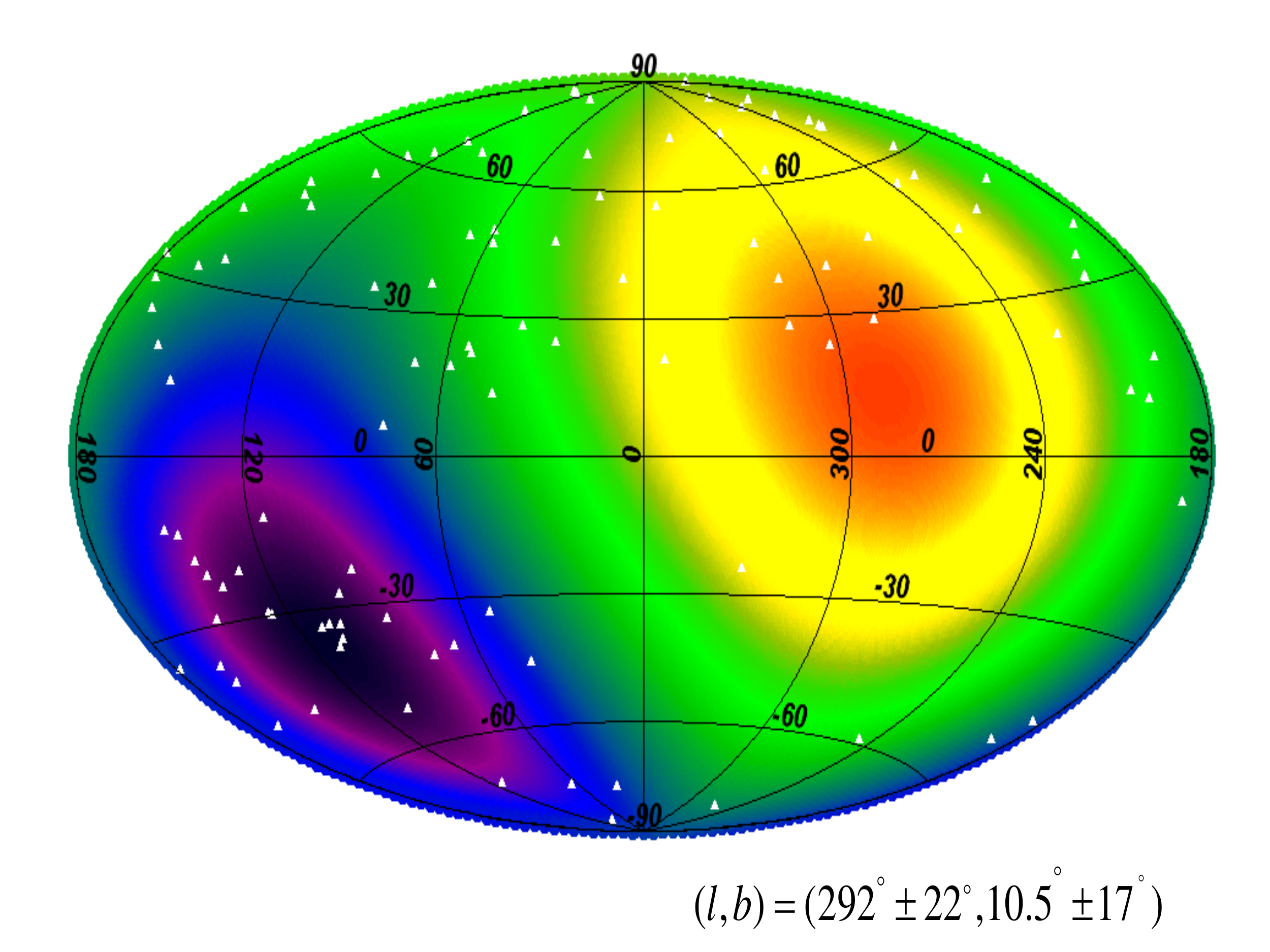}{0.5\textwidth}{(a)}
          \fig{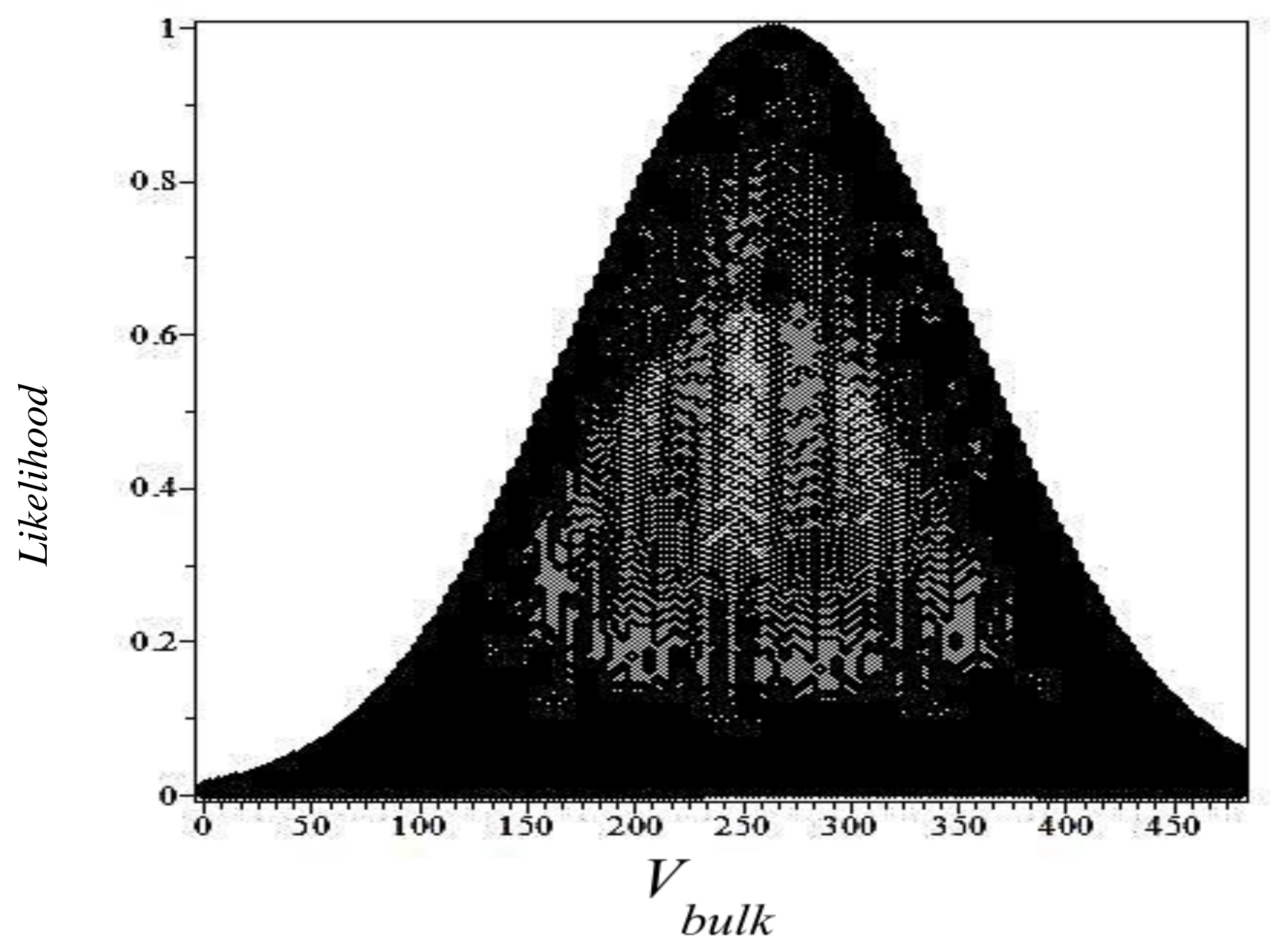}{0.5\textwidth}{(b)}
          }
\gridline{\fig{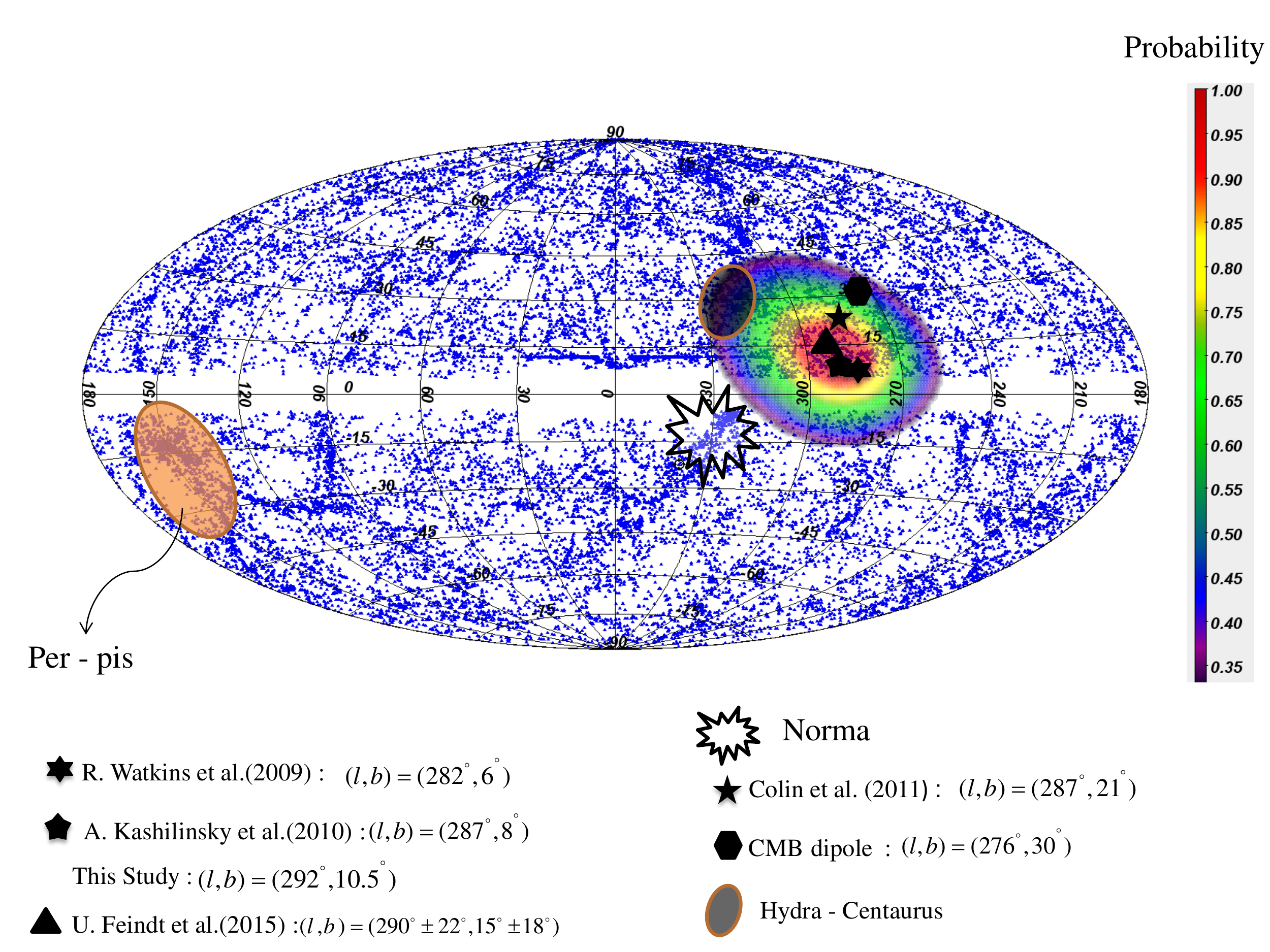}{0.97\textwidth}{(c)}
          }
\caption{\small{\emph{Probability of bulk flow direction in galactic longitude $l$ and galactic
latitude $b$ using $2\times 10^{5}$ datapoints for $0.15<z<0.035$.The most probable direction pointing towards$(l; b) = (292^{o}\pm22^{o}; 10.5^{o}\pm17^{o})$. Distribution of SNe Ia on the sky in galactic coordinates. In this Fig, the results of other studies have also been shown }}.\label{fig:p1}}
\end{figure*}

We first concentrate on the nearest redshift shell, $0.015 < z < 0.035$ $(45-105h^{-1} Mpc)$. This range includes 109 supernovas of 557 supernova Union2. We use the
maximum likelihood analysis method to find the bulk flow. Probability of bulk flow direction in galactic longitude $l$ and galactic
latitude $b$ using $2\times 10^{5}$ datapointss for $0.15<z<0.035$ have been shown in Fig(\ref{fig:p1}). As can be seen there is a bulk flow of \ \ $v_{bulk} =268^{+130}_{-130} kms^{-1}\ \  towards \ \ (l,b) =( 292^{o}\pm20^{o},10.5^{o}\pm17^{o})$.
In r.h.s of top panel of Fig(\ref{fig:p1}). the results of some studies which are comparable with our result at $(1-\sigma)$ confidence level have been shown.
Our results are very close to \citep{Colin} who found a bulk flow of $v_{bulk} = 250^{+190}_{-160} kms^{-1}$ towards $(l,b) =( 287^{o},21^{o})$
 ,\citep{Feindt} who estimate a bulk flow  of
$ v_{bulk} = 292^{+96}_{-96} kms^{-1}$ towards $(l,b) =( 290\pm22,15\pm18)$  and \citep{Wang} who found $ v_{bulk} = 271^{+101}_{-101} kms^{-1}$ towards $(l,b) =( 270\pm20,10\pm18)$ using the same data and in the same scale. Also the result is compatible with some previous studies at the same scale. Using a maximum likelihood approach, \citep{Watkins}computed
$  v_{bulk} = 416\pm78 kms^{-1}$towards$(l,b) =( 282^{o},6^{o})$.
Their results correspond to a sample with an effective Gaussian window of $50 h^{-1} Mpc$.
\citep{Hoffman} within a $60 h^{-1} Mpc$ top-hat sphere, based on the Mark III peculiar velocity catalogue found
$  v_{bulk} = 366\pm78 kms^{-1}\ \ towards \ \ (l,b) =( 300^{o},13^{o})$
At this scale , \citep{Turnbull} presented new ‘minimal variance’ bulk flow measurements based
upon the ‘First Amendment’ compilation of 245 Type Ia supernovae (SNe) peculiar velocities
and find a bulk flow of $ v_{bulk}=249 \pm 76 kms^{-1} towards  (l,b) =( 319^{o},7^{o})$
For a sphere of radius $40h^{-1} Mpc$ centered on the MW, \citep{Nusser} derive a bulk flow of
$ v_{bulk}=333 \pm 38 kms^{-1}$ towards $ (l,b) =( 276^{o},14^{o})$.
 While at this scale the direction of bulk motion in our study is consistent with\citep{Kashlinsky2} who found
$ v_{bulk} \simeq 1000 kms^{-1}\ \ towards \ \ (l,b) =( 287^{o},7^{o})$
at $(1\sigma)$, the amplitude is much lower and aligned with expectation of $\Lambda CDM$.
 Using a statistical method based on an optimized cross-correlation with nearby
galaxies, \citep{Lavaux2} extract the kSZ signal generated by plasma halo of galaxies from the cosmic microwave background (CMB) temperature anisotropies observed by the Wilkinson Microwave
Anisotropy Probe (WMAP). By considering only the galaxies within $ 50h^{-1}Mpc$ they found
$ v_{bulk} \simeq 533\pm263 kms^{-1}\ \ towards \ \ (l,b) =( 324\pm27^{o},-7\pm17^{o})$.
 Although, we find that the direction of bulk motion at this scale is
 approximately aligned with the direction of the CMB dipole and Hydra–Centaurus supercluster at $(1\sigma)$ confidence level, the amplitude of bulk flow is less than half of the amplitude of the CMB dipole.

\subsection{Numerical analysis for redshift $0.015<z<0.06$}
\begin{figure*}
\gridline{\fig{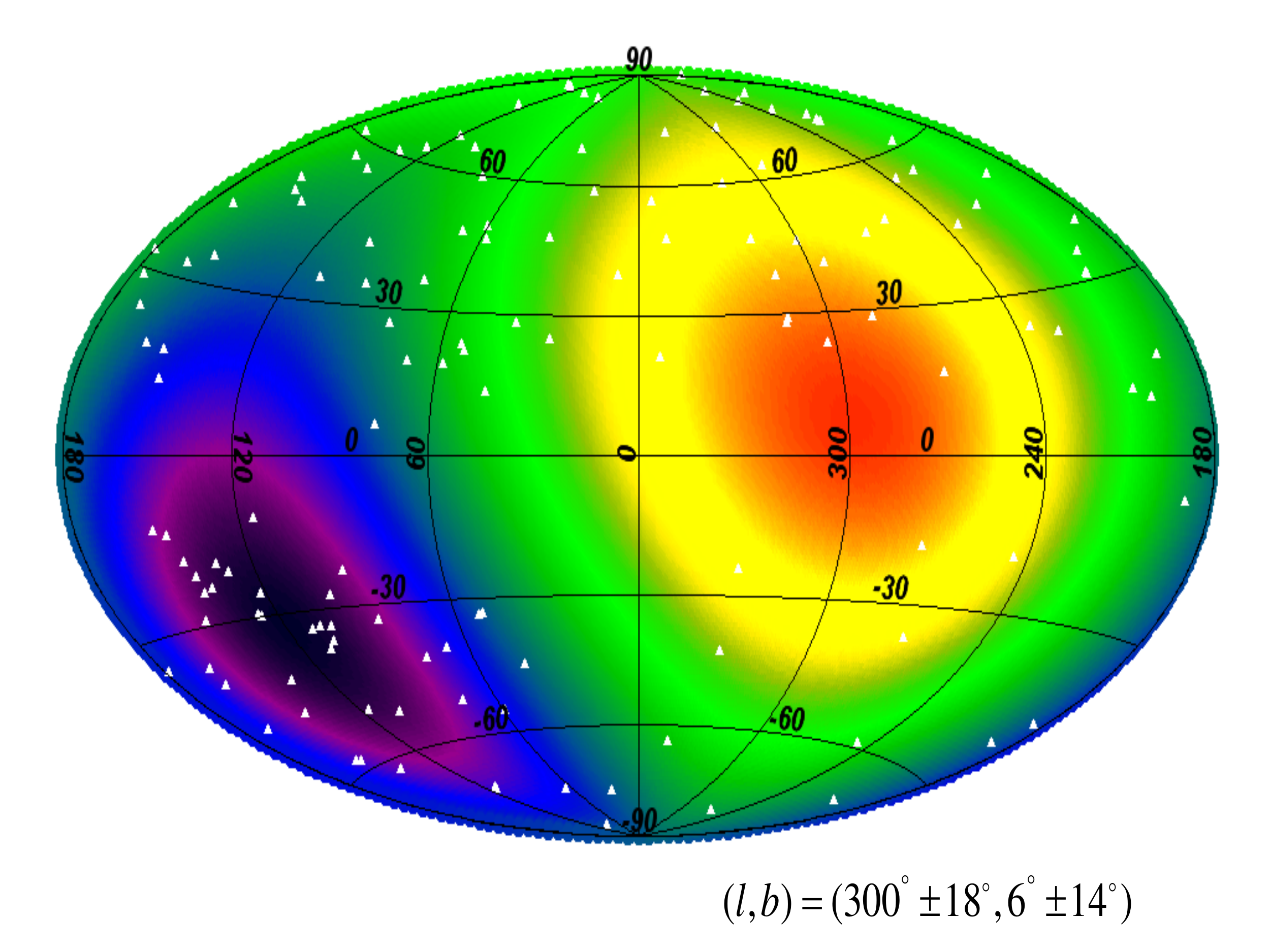}{0.5\textwidth}{(a)}
          \fig{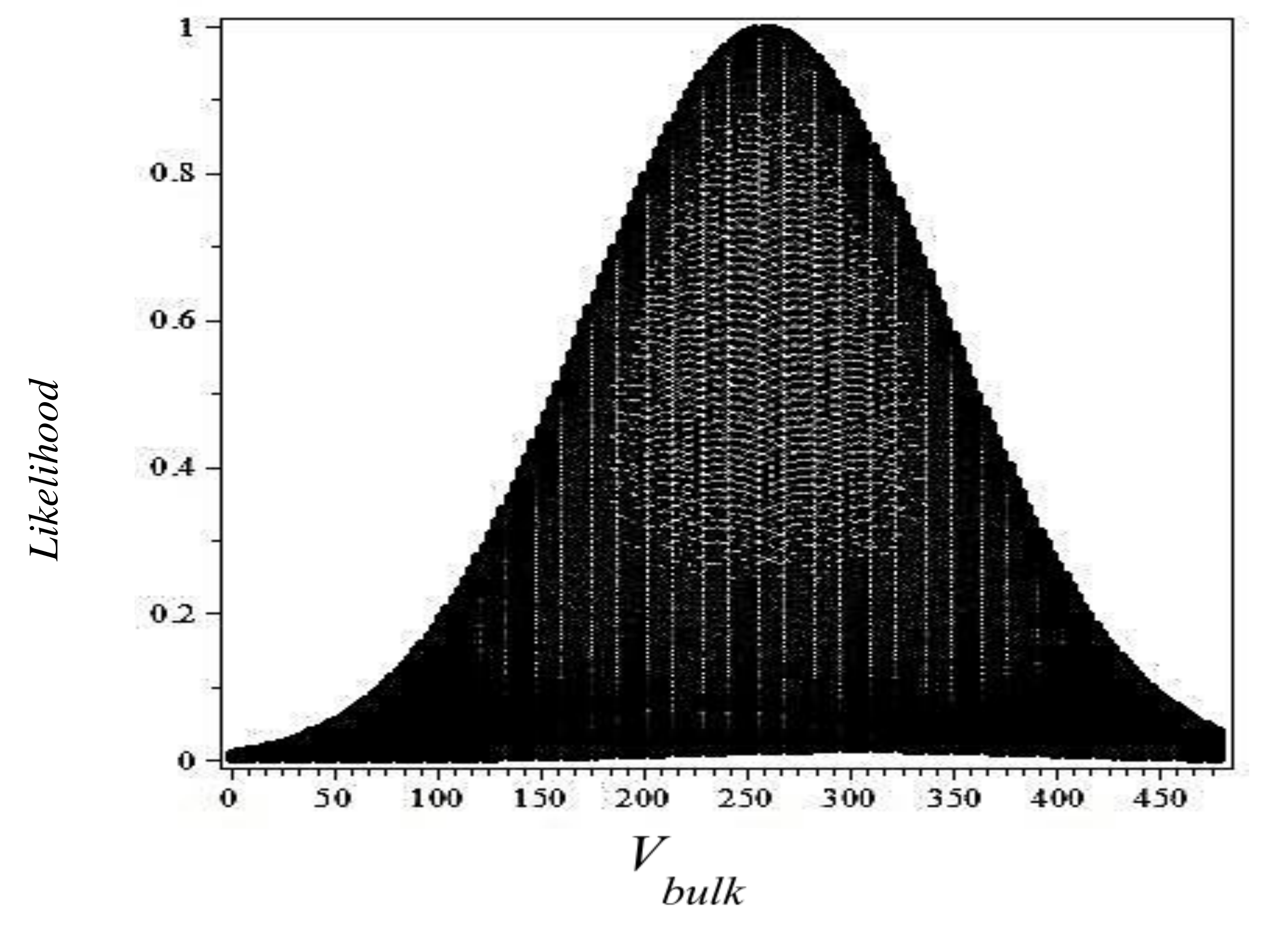}{0.5\textwidth}{(b)}
          }
\gridline{\fig{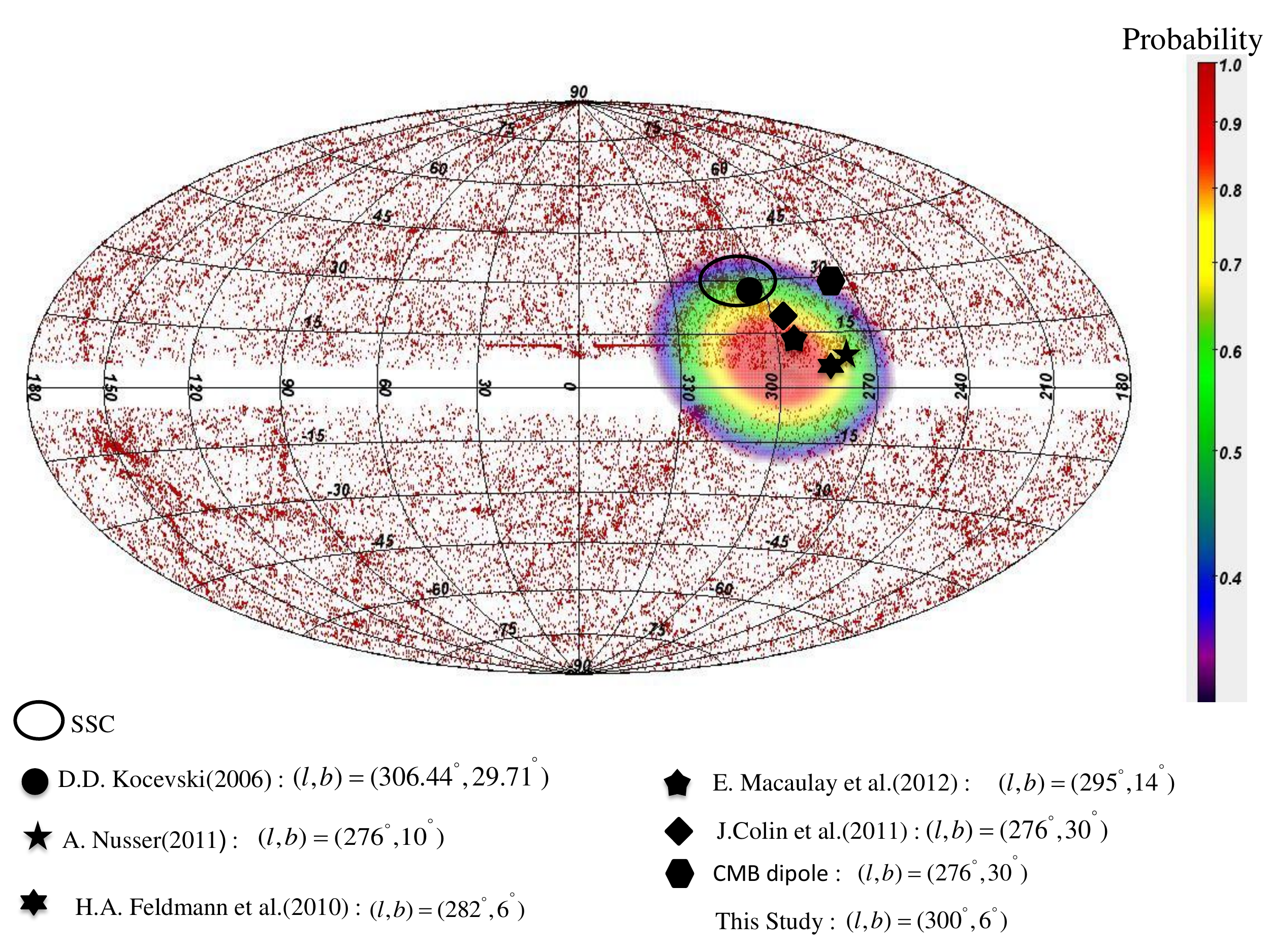}{0.97\textwidth}{(c)}
          }
\caption{\small{\emph{Probability of bulk flow direction in galactic longitude $l$ and galactic
latitude $b$ using $2\times 10^{5}$ datapoints for $0.15<z<0.06$.The most probable direction pointing towards$(l; b) = (300^{o}\pm18^{o}; 6^{o}\pm14^{o})$. Distribution of SNe Ia on the sky in galactic coordinates. The result of other studies have also been shown}}.\label{fig:p2}}
\end{figure*}

Such as \citep{Lavaux}, we find that less than half of the amplitude of the CMB dipole is generated within a volume enclosing the
Hydra–Centaurus–Norma super cluster at around $40 h^{-1} Mpc$.
\citep{Kocevski} found that the GA only accounts for 44\% of the dipole anisotropy in a large X-ray cluster sample, with the rest evidently caused by more distant sources such as the Shapley Supercluster (SSC) at a distance of [$105-165h^{-1}Mpc$] $(0.035 < z < 0.055)$ in the direction $ (l, b) = (306.44^{o}, 29.71^{o})$.
 Due to dominant superclusters such as Shapley Supercluster (SSC) at a distance of $105-165h^{-1}Mpc$ $(0.035 < z < 0.055)$, it is believed that it be largely responsible for this bulk flow. Hence most of the studies have been focused in this region. However, it is expected that both Hydra–Centaurus–Norma super cluster and Shapley Supercluster (SSC) affect the motion of Local Group. Hence, in order to consider the both effects together, we perform our analysis in $(0.015 < z < 0.06)$ region. This range includes 142 supernova of 557 supernovas Union2. Fig(\ref{fig:p2}) shows the results of our analysis. As can be seen , we find the bulk flow of
 $v_{bulk} \simeq 257\pm120 kms^{-1}$\ \ towards \ \ $(l,b) =( 300^{o}\pm18^{o},6^{o}\pm14^{o})$
This direction is very close to centaurus constellation and aligned with direction of (SSC) at $(1-\sigma)$ confidence level, however bulk flow and  direction of CMB dipole does not improve. Our results are consistent with some previous studies.
\citep{Kocevski}; \citep{Nusser}; \citep{Feldman}; \citep{Macaulay}; \citep{Colin}.

\subsection{Numerical analysis for redshift $0.015<z<0.1$}
\begin{figure*}
\gridline{\fig{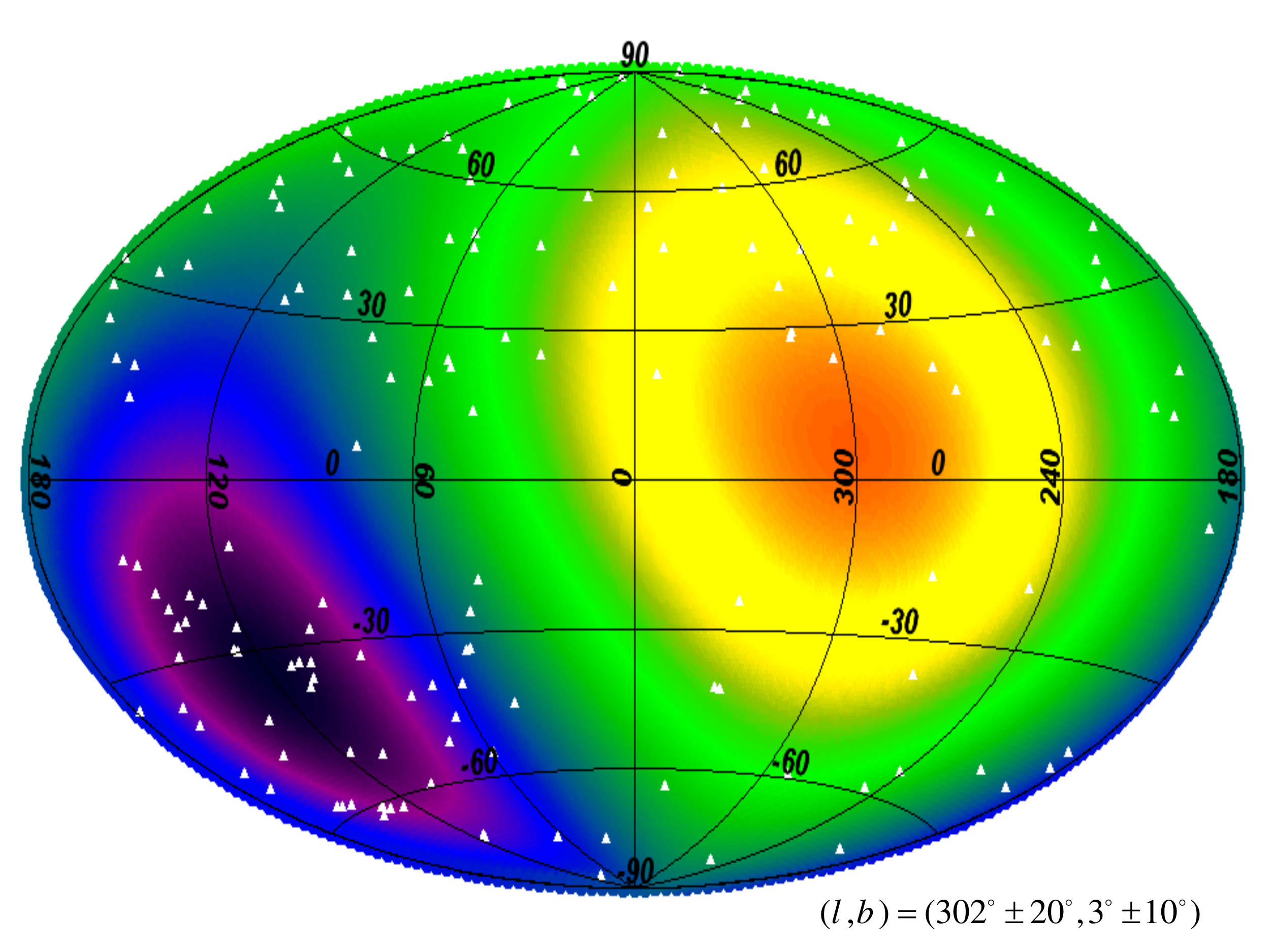}{0.5\textwidth}{(a)}
          \fig{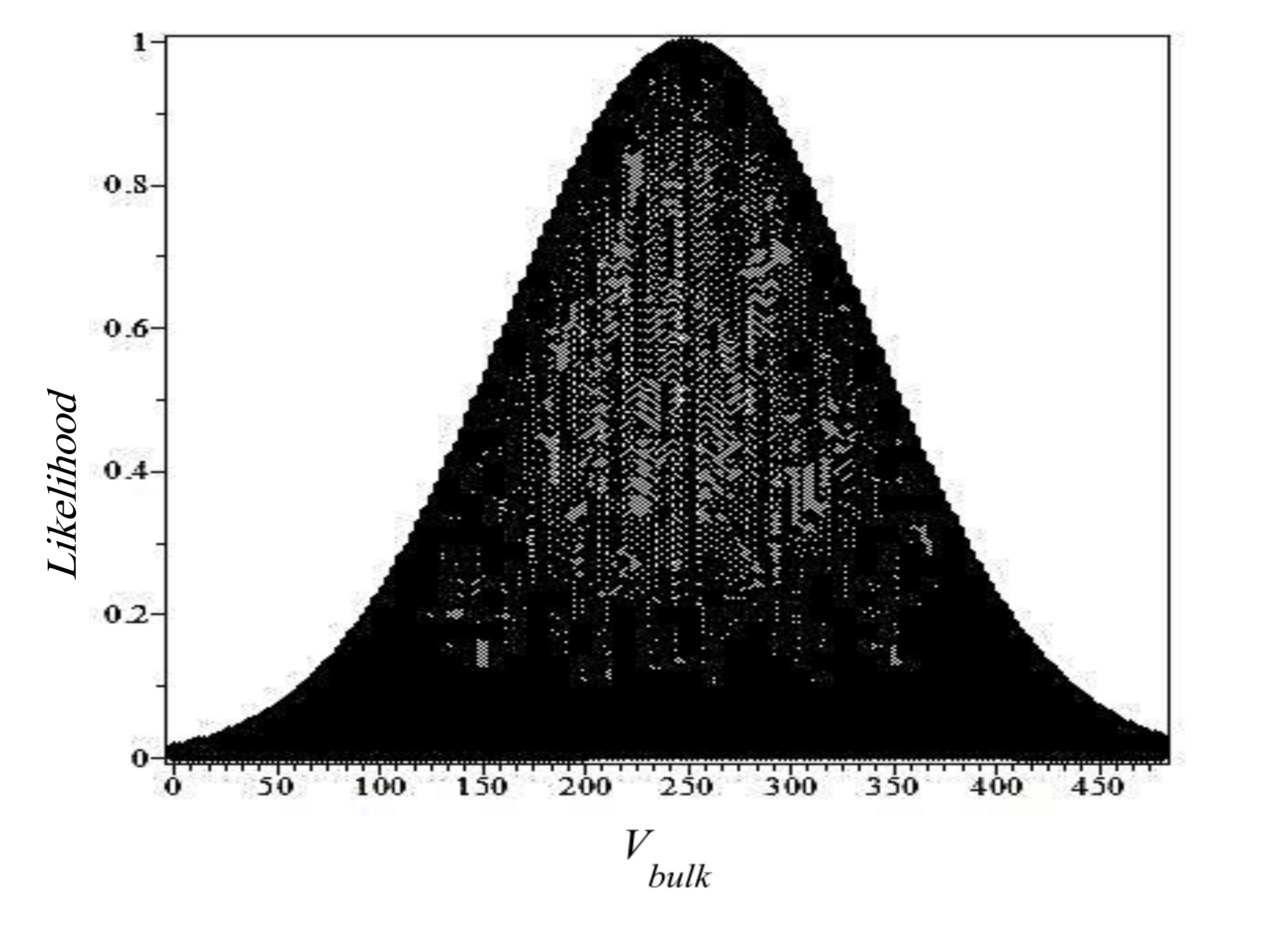}{0.5\textwidth}{(b)}
          }
\gridline{\fig{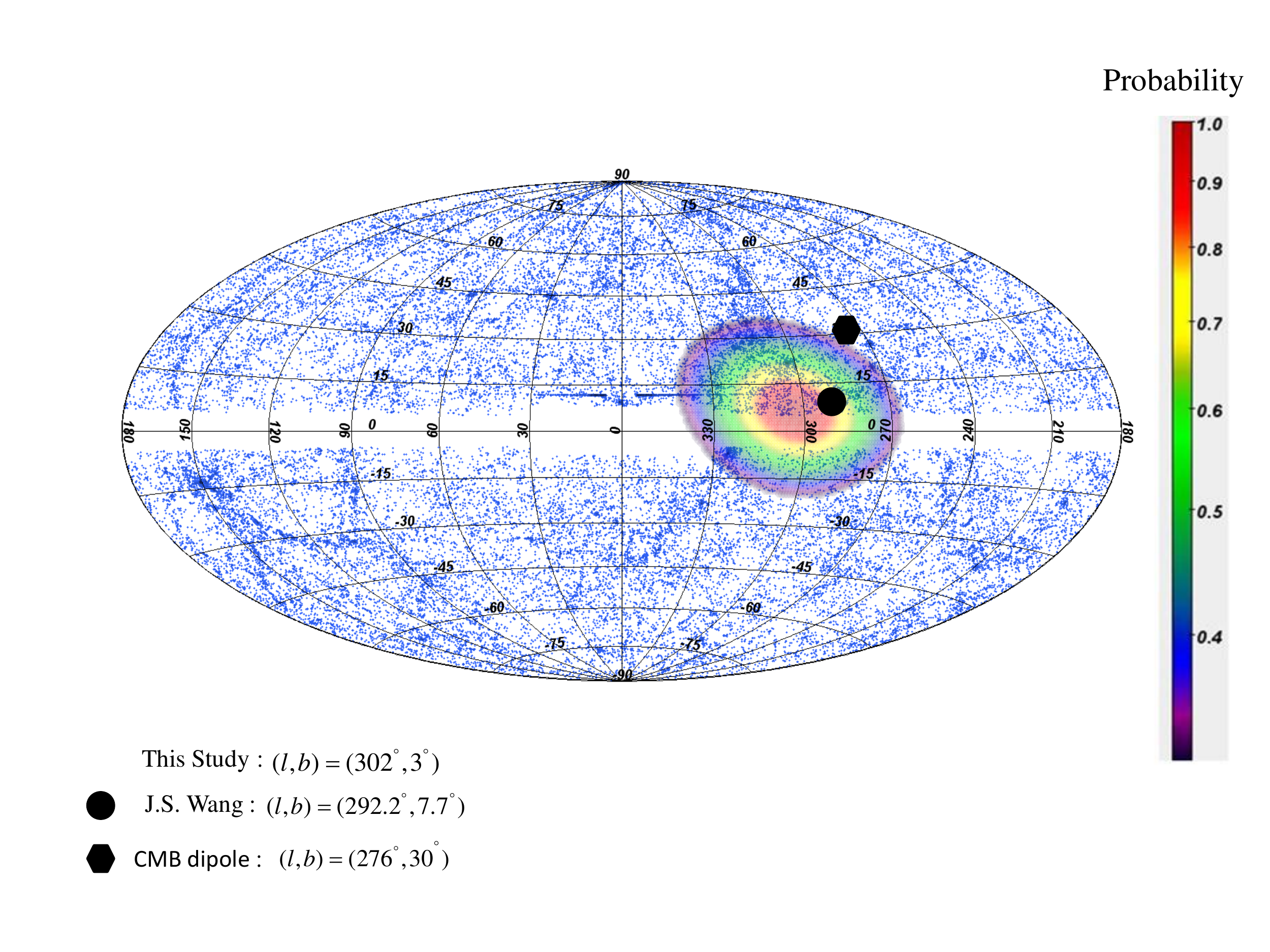}{0.97\textwidth}{(c)}
          }
\caption{\small{\emph{Probability of bulk flow direction in galactic longitude $l$ and galactic
latitude $b$ using $2\times 10^{5}$ datapoints for $0.15<z<0.1$.The most probable direction pointing towards$(l; b) = (302^{o}\pm20^{o}; 3^{o}\pm10^{o})$. Distribution of SNe Ia on the sky in galactic coordinates. The results of other studies have also been shown}}.\label{fig:p3}}
\end{figure*}

 A large number of authors suggest that one has to go
at least as far as the Shapley concentration at about $150h^{-1}Mpc$
in order to fully recover the dipole motion \citep{Kocevski}; \citep{Hoffman}; \citep{Lavaux}; \citep{Shapley}; \citep{Scaramella}; \citep{Raychaudhury}. The tentative observation show that the dipole motion does not appear
to converge at a distances scale of the SSC,\ \ i.e. $150h^{-1}Mpc$ and convergence
 must occur well beyond $(z > 0.06)$ \citep{Colin}. Due to dominant superclusters such as Shapley or Horologium–Reticulum in the southern
hemisphere at scales above $120 h^{-1} Mpc$, one might need to go well beyond $200 h^{-1} Mpc$ to fully recover the
dipole vector \citep{Watkins}. Here we make analysis for redshift range $0.015<z<0.1$.
This range includes 165 supernovas of 557 supernova Union2. Fig(\ref{fig:p3}) shows the results of our analysis. As can be seen , we find the bulk flow of
 $v_{bulk} \simeq 257\pm120 kms^{-1}$\ \ towards \ \ $(l,b) =( 302^{o}\pm20^{o},3^{o}\pm10^{o})$
\citep{Wang} find a dipolar anisotropy in the direction $(l,b)= ( 309.2 \pm 15.8^{o},8.6 \pm 10.5^{o})$ in galactic coordinates with a significant evidence 97.29\%
(more than $2 \sigma$). The direction and velocity of redshift range $0.015 < z < 0.1$ are consistent with the
results from $0.015 < z < 0.035$ and $0.015<z< 0.06$. The consistency between the results of high and low redshift may be interpreted as the following possibilities.\\\\
$\bullet$ In addition to attraction due to nearby over densities, the anisotropy may be caused by the other effects such as dark energy dipole, hence due to the non-local effect of dark energy, the direction is constant on all cosmic scale. If the anisotropy is caused only by the peculiar velocity, the anisotropic direction should be randomly distributed on different cosmic scales, because peculiar velocity is driven by emergent of large scale structure. \citep{Cai}\\
$\bullet$  Because of sparseness of the data at high redshift, the high-redshift results may be contaminated by the low redshift data.\\
So redshift tomography method may tell the differences between the dark energy dipole and peculiar velocity if high-z SNIa data are available \citep{Cai}

\subsection{Numerical analysis for redshift $0.015<z<1.4$}
\begin{figure*}
\gridline{\fig{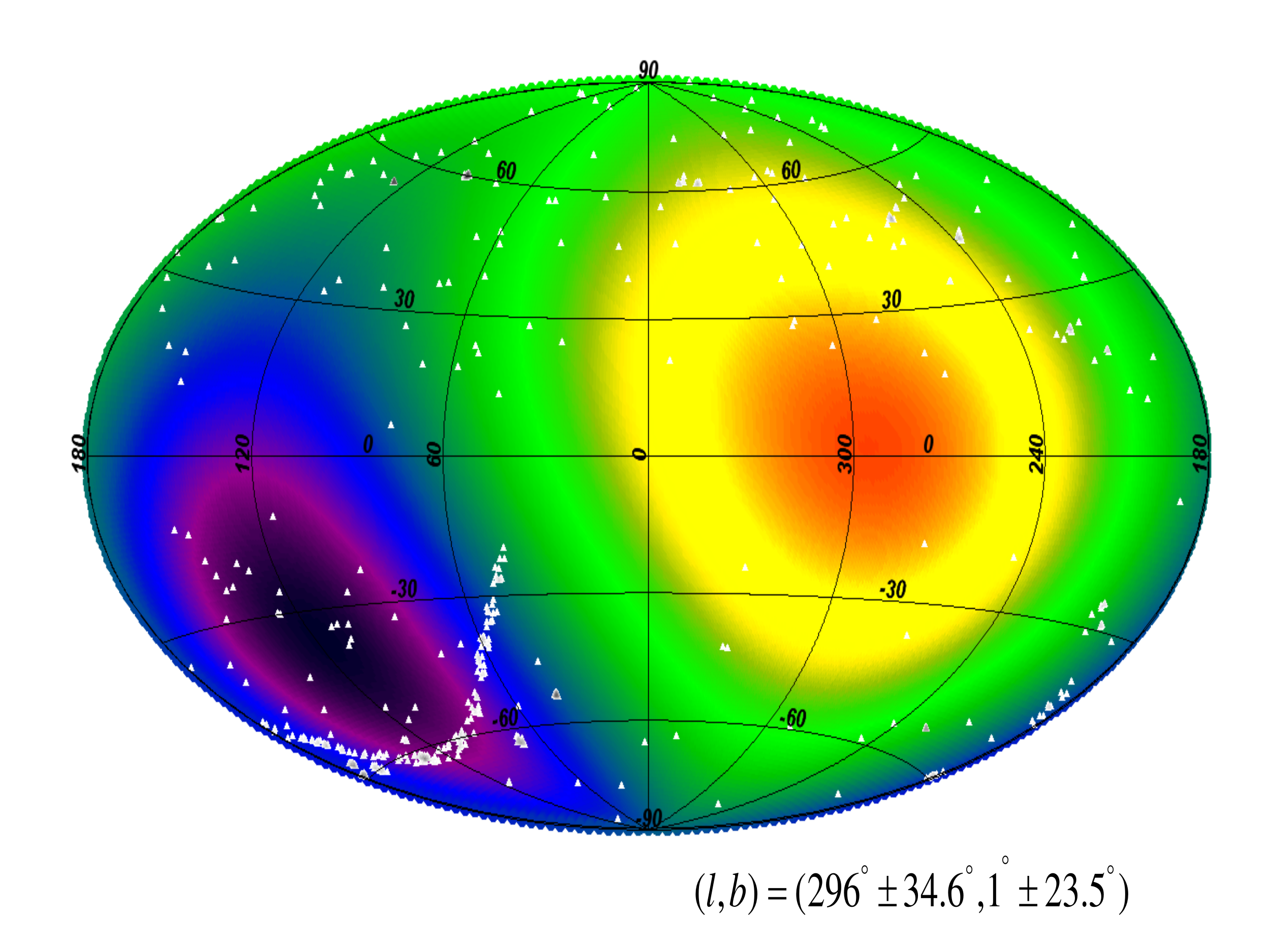}{0.5\textwidth}{(a)}
          \fig{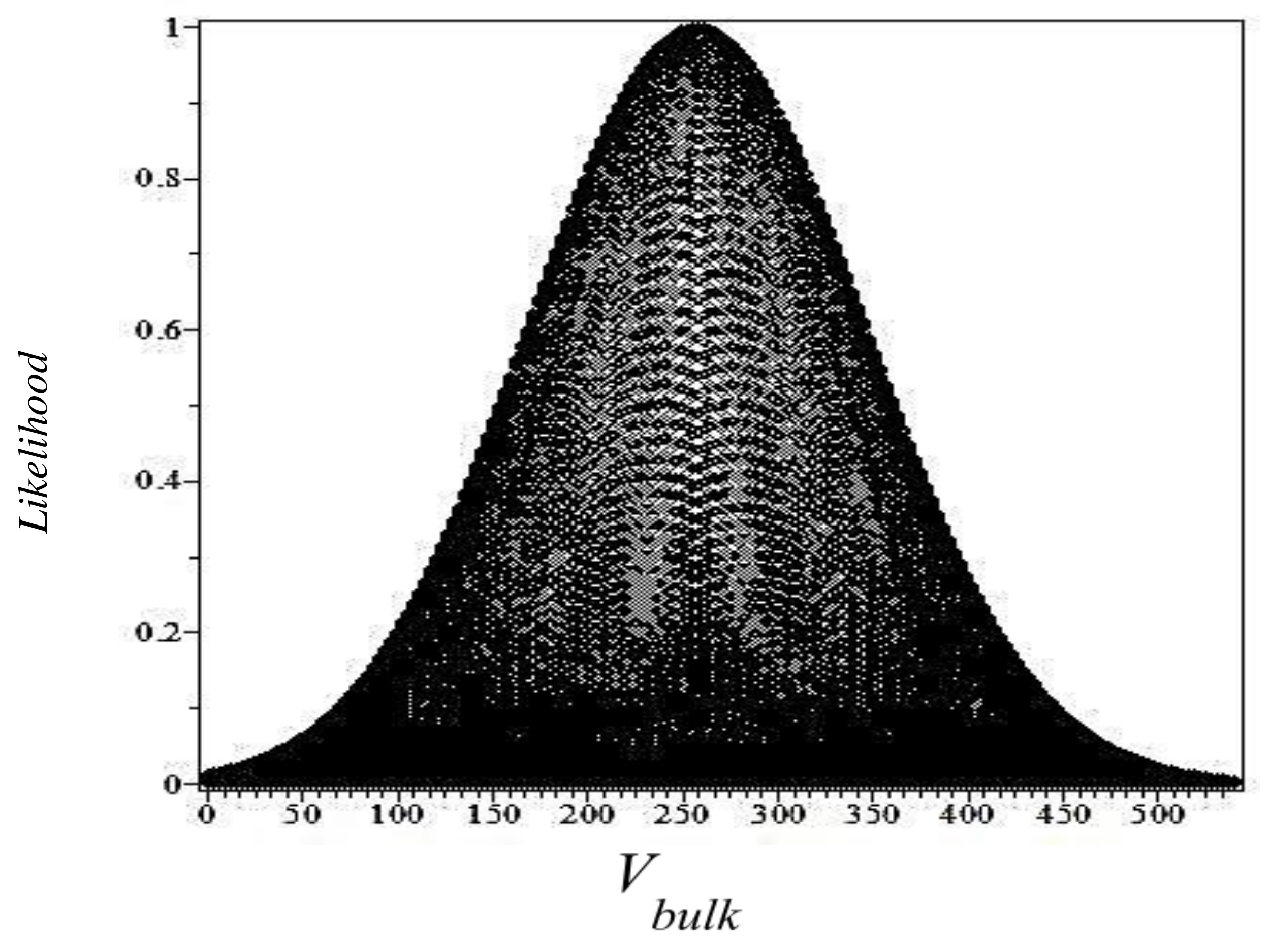}{0.5\textwidth}{(b)}
          }
\gridline{\fig{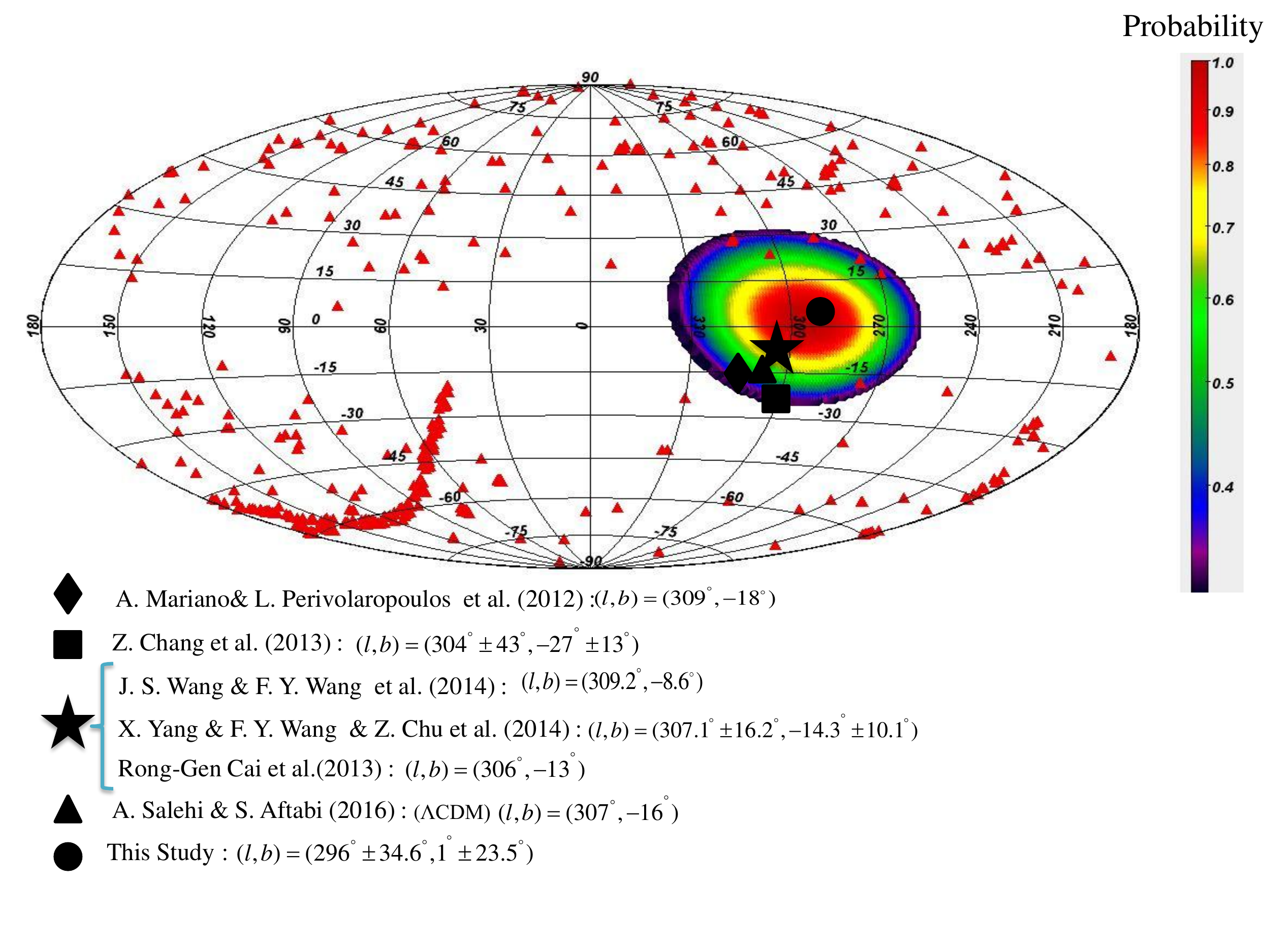}{0.97\textwidth}{(c)}
          }
\caption{\small{\emph{Probability of bulk flow direction in galactic longitude $l$ and galactic
latitude $b$ using $2\times 10^{5}$ datapoints for $0.015<z<1.4$.The most probable direction pointing towards$(l; b) = (296^{o}\pm34^{o}; 1^{o}\pm23.5^{o})$. Distribution of SNe Ia on the sky in galactic coordinates. Red triangulares denote SNe with $0.015<z<1.4$. The results of other studies have also been shown }}.\label{fig:p4}}
\end{figure*}

\begin{figure*}
\gridline{\fig{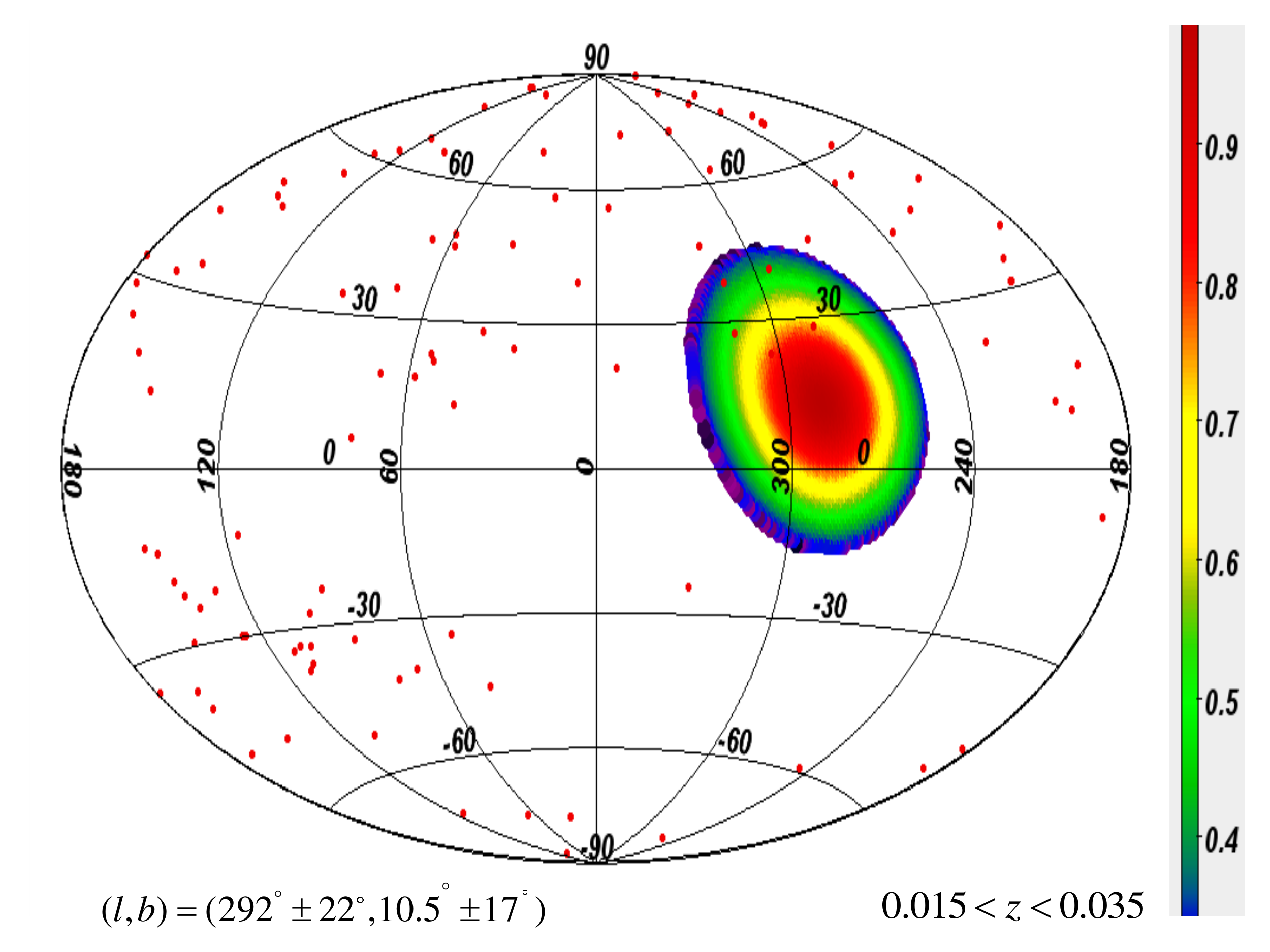}{0.5\textwidth}{(a)}
          \fig{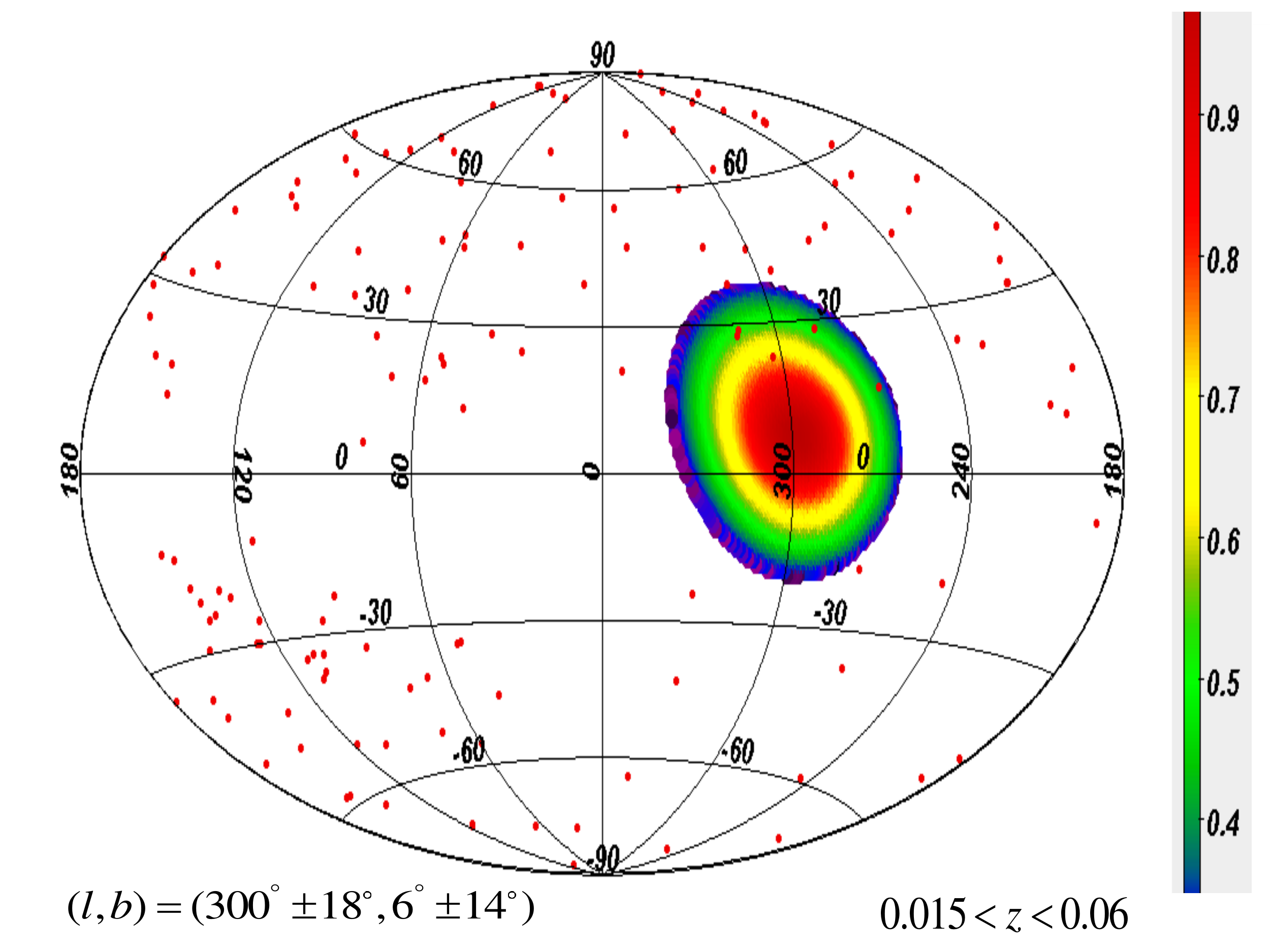}{0.5\textwidth}{(b)}
          }
\gridline{\fig{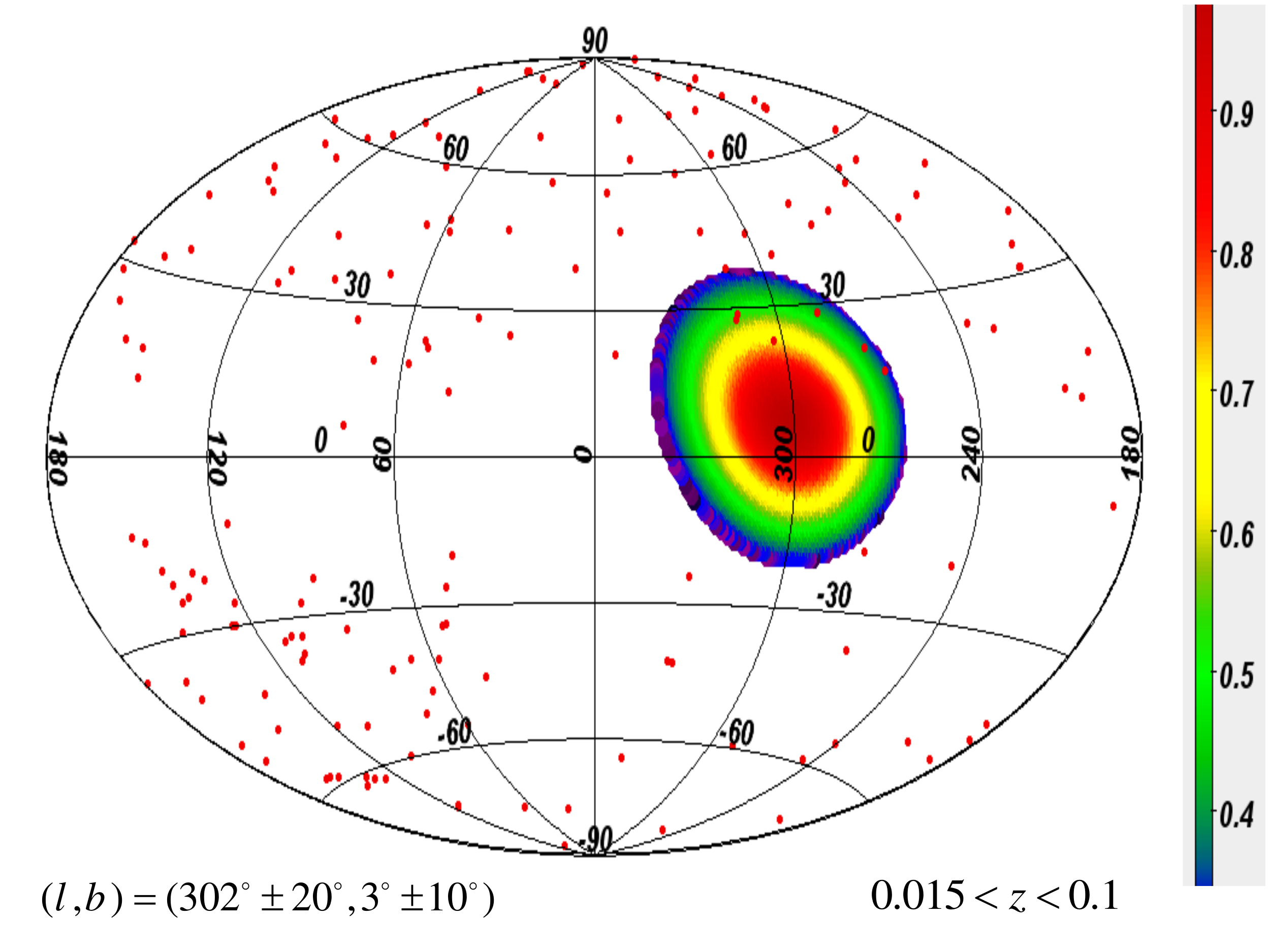}{0.5\textwidth}{(c)}
          \fig{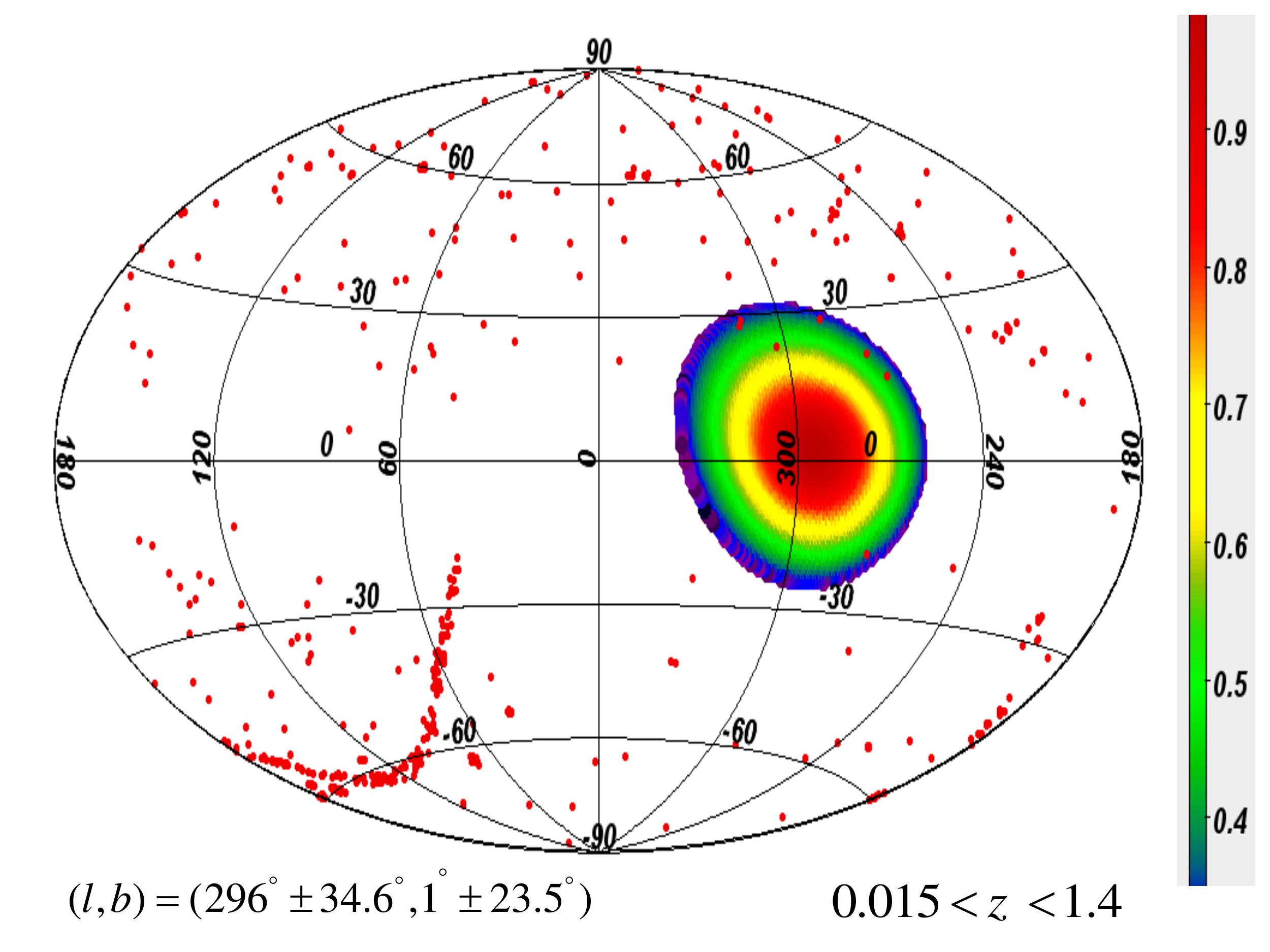}{0.5\textwidth}{(d)}
          }
\caption{\emph{The plot of ($1\sigma$) confidence level of bulk flow direction in galactic longitude $l$ and galactic
latitude $b$ for redshift ranges}; $0.015<z<0.035$, $0.015<z<0.06$, $0.015<z<0.1$ and  $0.015<z<1.4$.\label{fig:p5}}
\end{figure*}

\begin{figure*}
\gridline{\fig{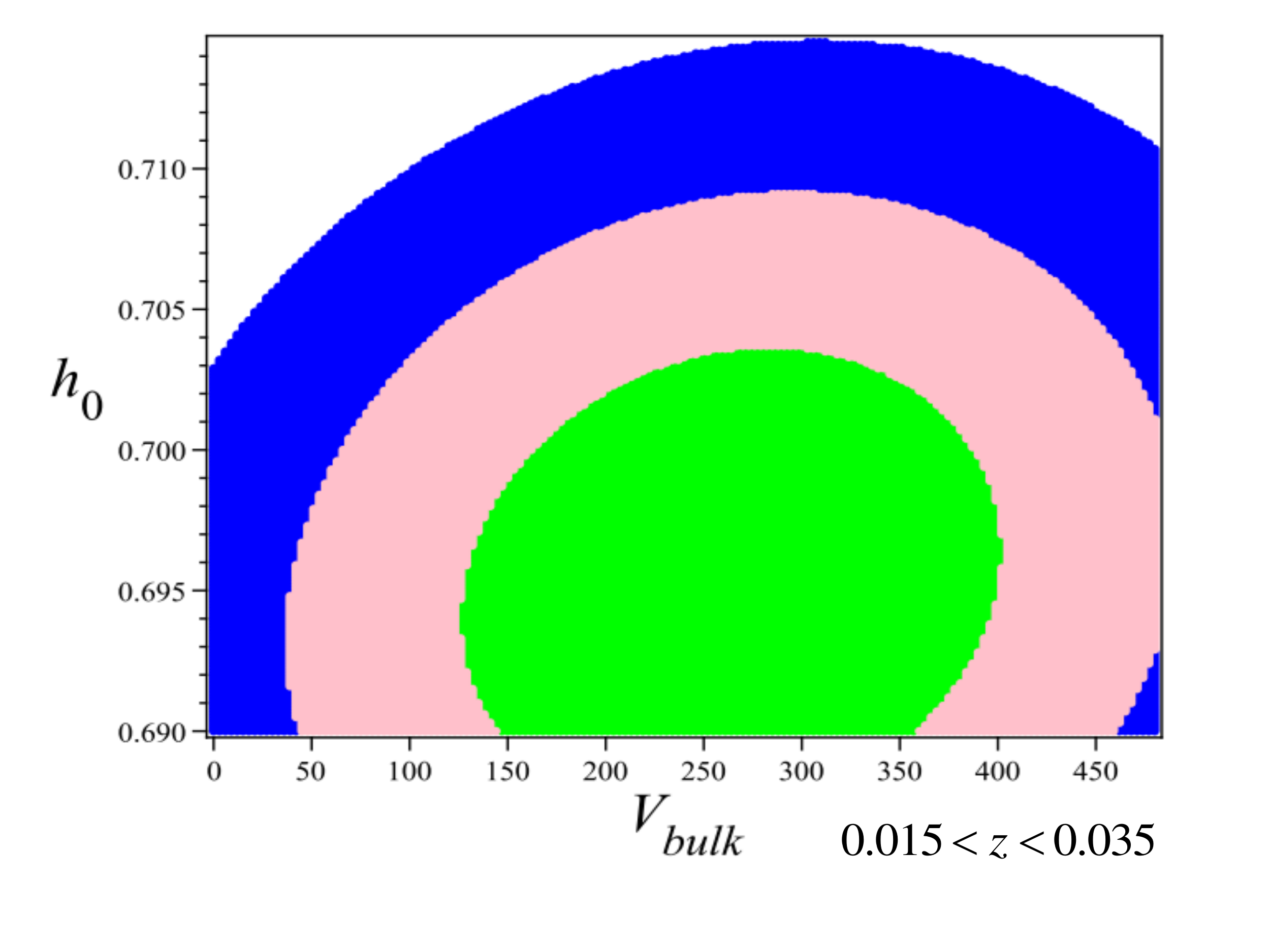}{0.5\textwidth}{(a)}
          \fig{Doc17-eps-converted-to.pdf}{0.5\textwidth}{(b)}
          }
\gridline{\fig{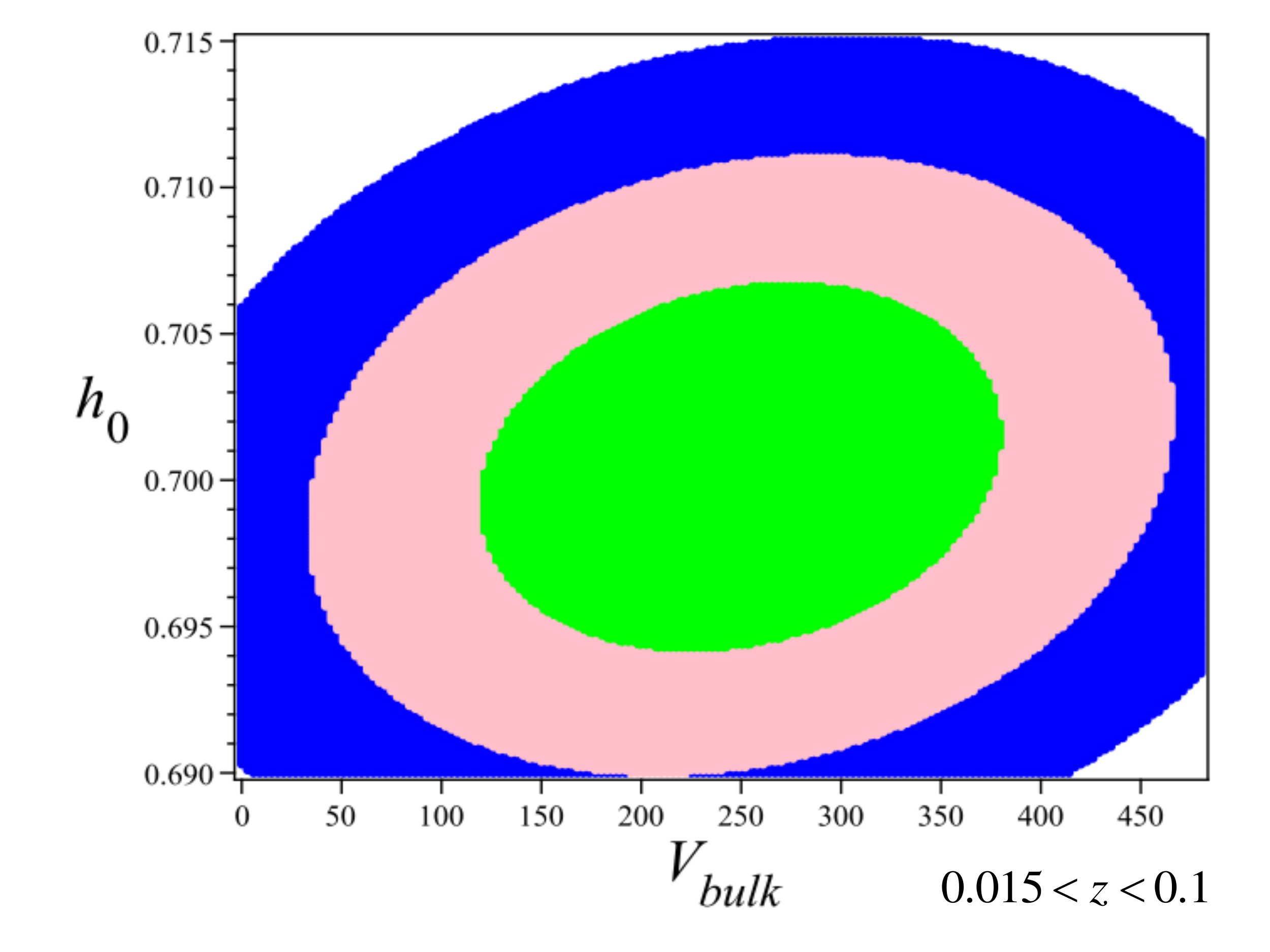}{0.5\textwidth}{(c)}
          \fig{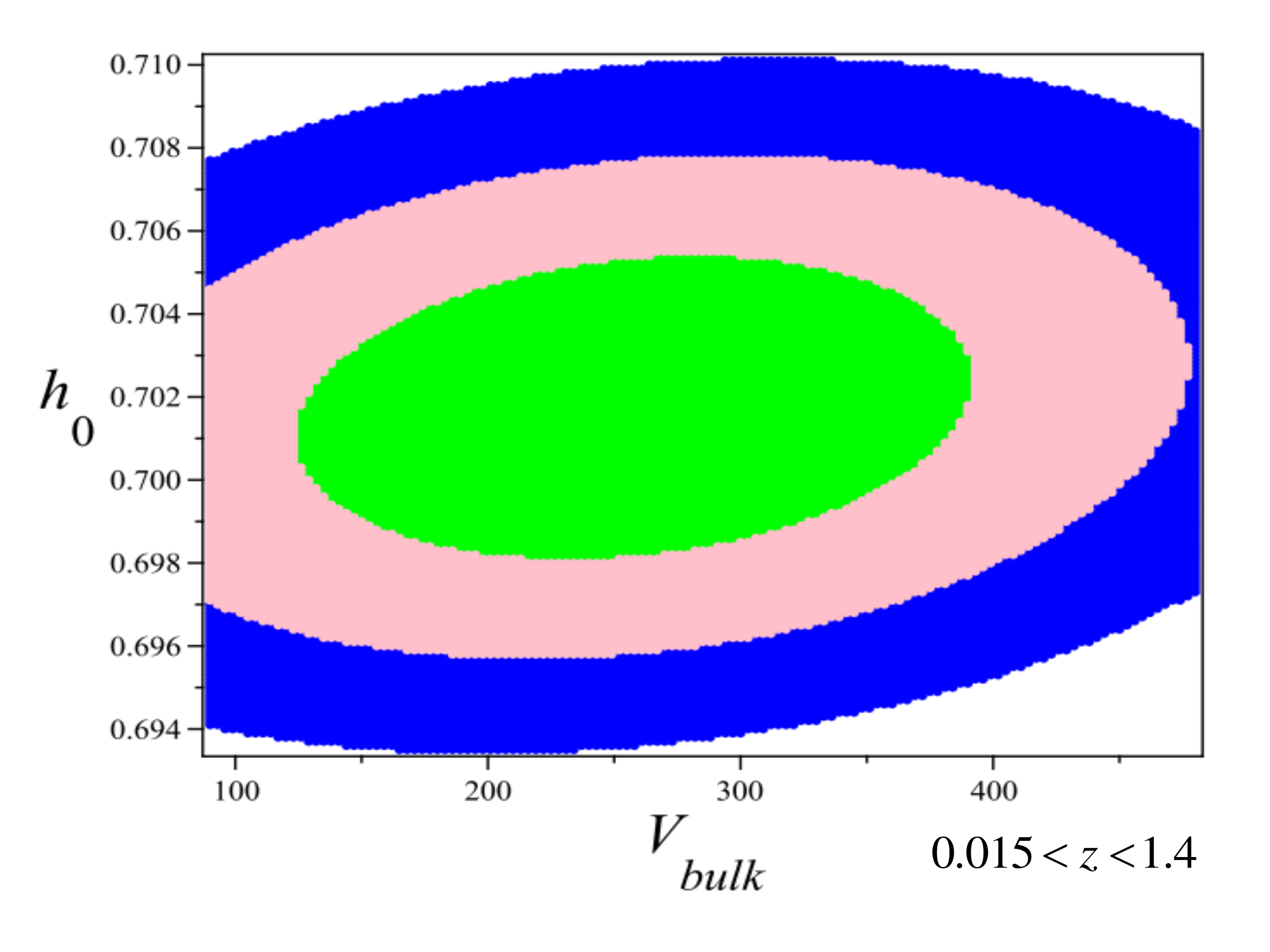}{0.5\textwidth}{(d)}
          }
          \caption{\emph{The plot of ($1,2,3\sigma$) confidence level of ($v_{bulk},h_{0}$}) for redshift ranges; $0.015<z<0.035$, $0.015<z<0.06$, $0.015<z<0.1$ and  $0.015<z<1.4$.\label{fig:p6}}
\end{figure*}

Here we use full union data to test the isotropy of the universe.(we find the bulk flow of
 $v_{bulk} \simeq 253 kms^{-1}$\ \ towards \ \ $(l,b) =( 296^{o}\pm34.6^{o},1^{o}\pm23.5^{o})$) in galactic coordinates. The result is compatible with the results of pervious studies of dark energy dipole  in this redshift, \citep{Mariano}; \citep{Chang}; \citep{Wang}; \citep{Yang};\citep{Cai}; \citep{Salehi}(see Fig \ref{fig:p4}). As an interesting result, the direction, magnitude of bulk flow and $h_{0}$ are approximately the same for all slices which contain low resift range $0.015<z<0.035$. Also their ($1-\sigma$) errors are compatible (see Fig (\ref{fig:p5}) and (\ref{fig:p6})). This hints that the high-redshift results may be contaminated by the lowred shift data, hence it encourages us to perform a ‘cosmic tomography’ in which the data are sliced up in redshift and the question of isotropy is studied separately
for each slice.

\begin{figure*}
\gridline{\fig{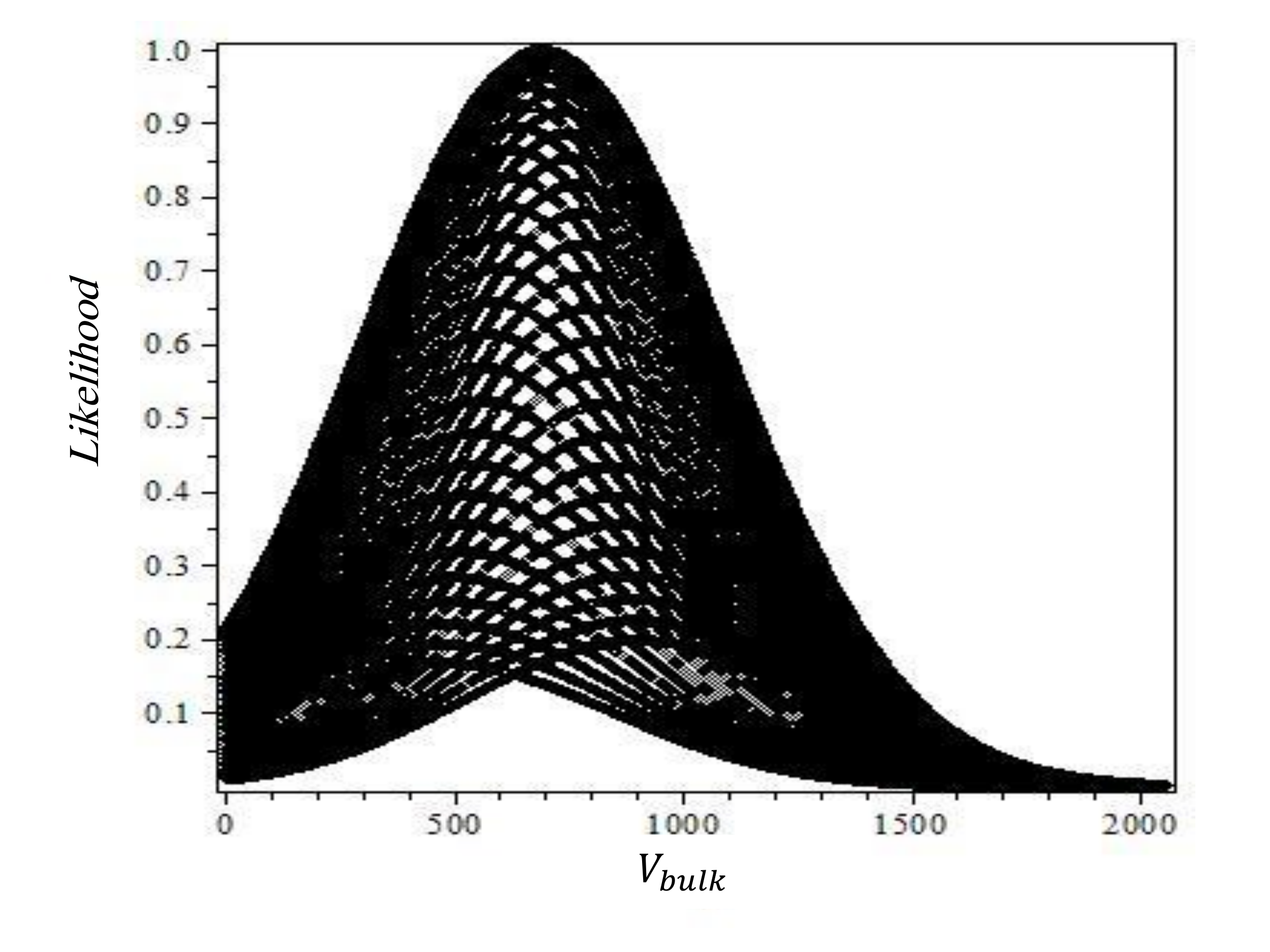}{0.5\textwidth}{(a)}
          \fig{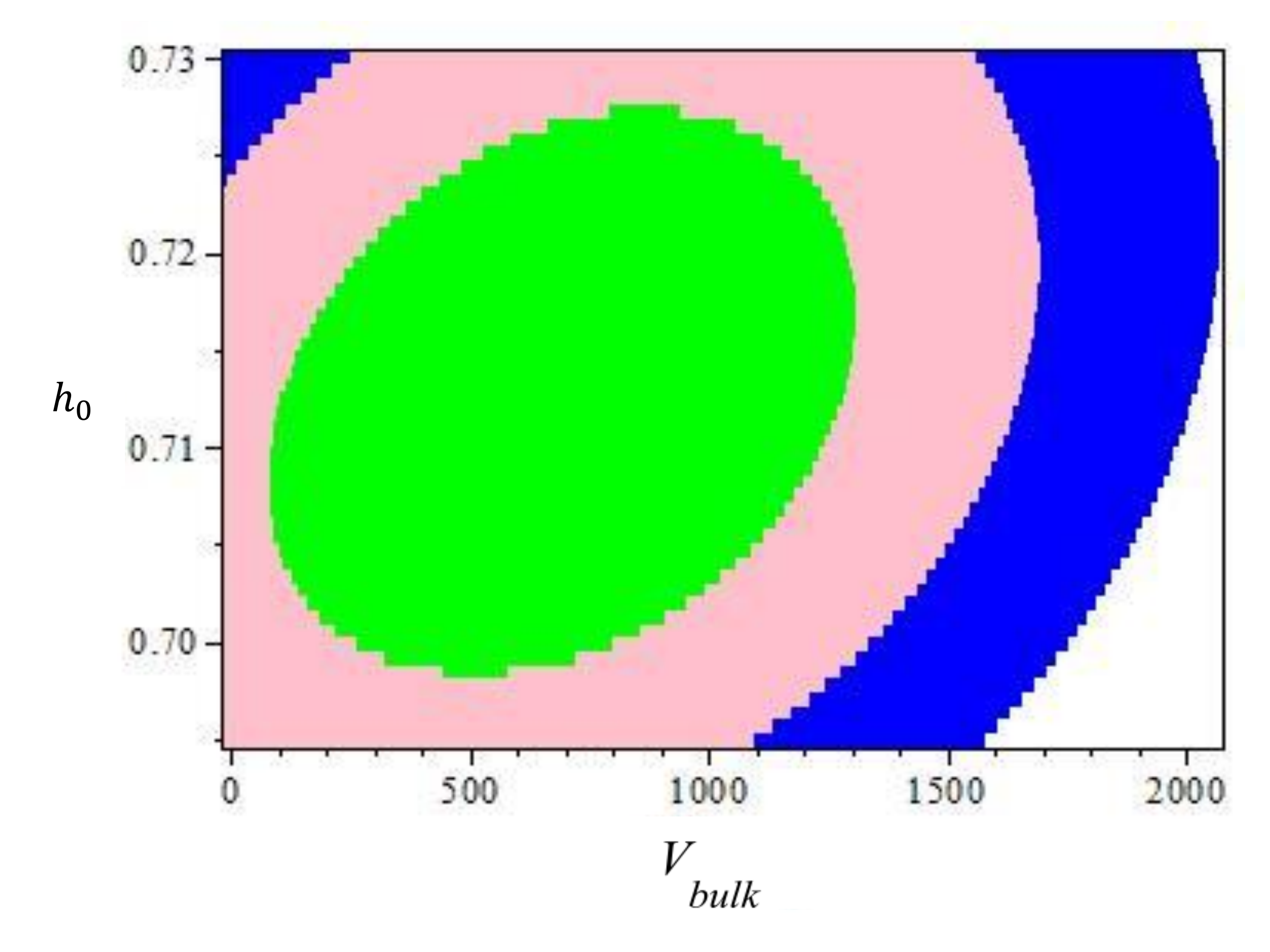}{0.5\textwidth}{(b)}
          }
\gridline{\fig{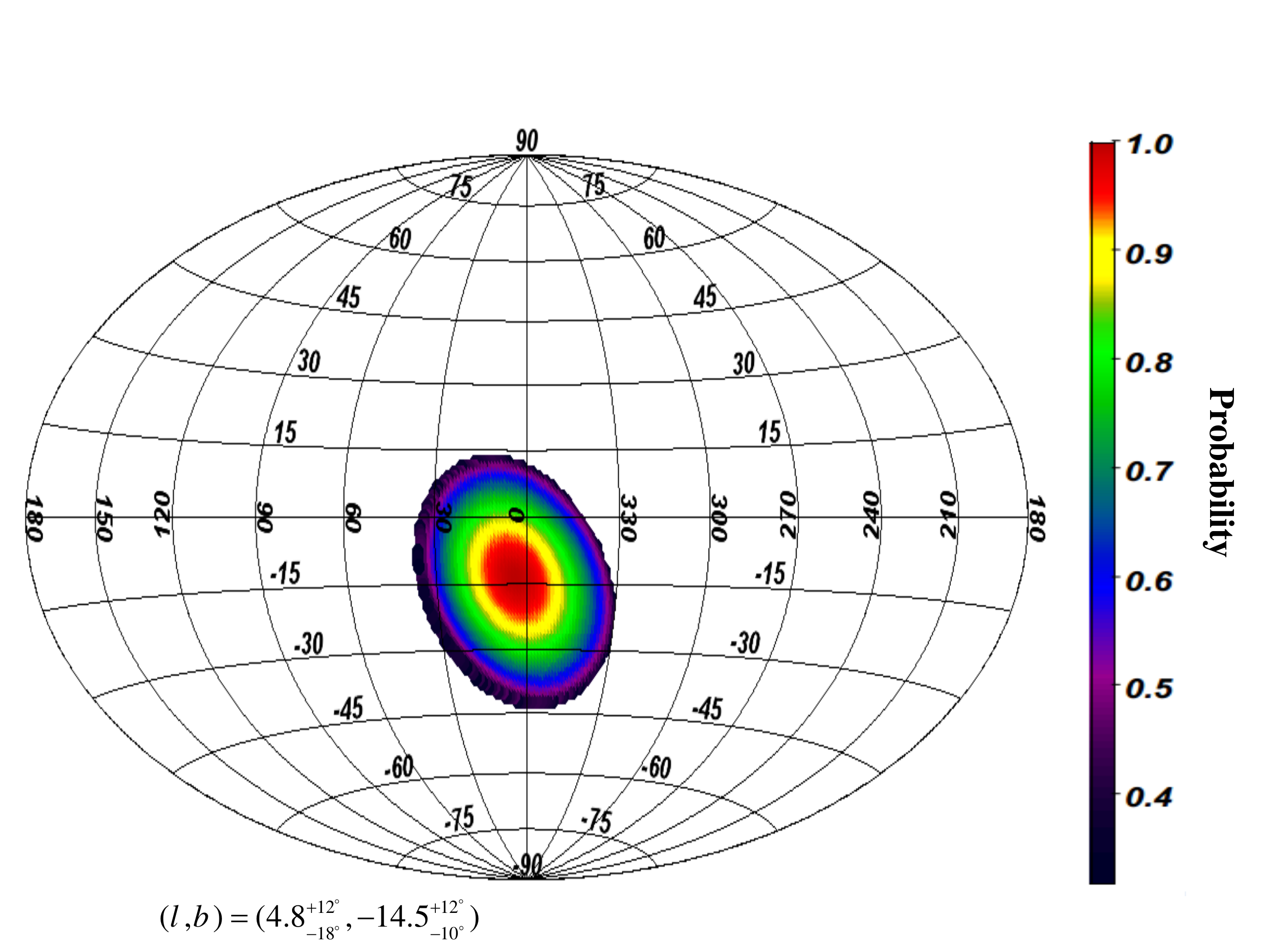}{0.97\textwidth}{(c)}
          }
\caption{\small{\emph{Probability of bulk flow direction in galactic longitude $l$ and galactic
latitude $b$ for $0.035<z<0.06$.The most probable direction pointing towards$(l; b) = (4.8^{+12^{o}}_{-18^{o}}; -14.5^{+12^{o}}_{-10^{o}})$. Distribution of SNe Ia on the sky in galactic coordinates. Red triangulares denote SNe with $0.035<z<0.06$. The results of other studies have also been shown }}.\label{fig:p7}}
\end{figure*}
\begin{figure*}
\gridline{\fig{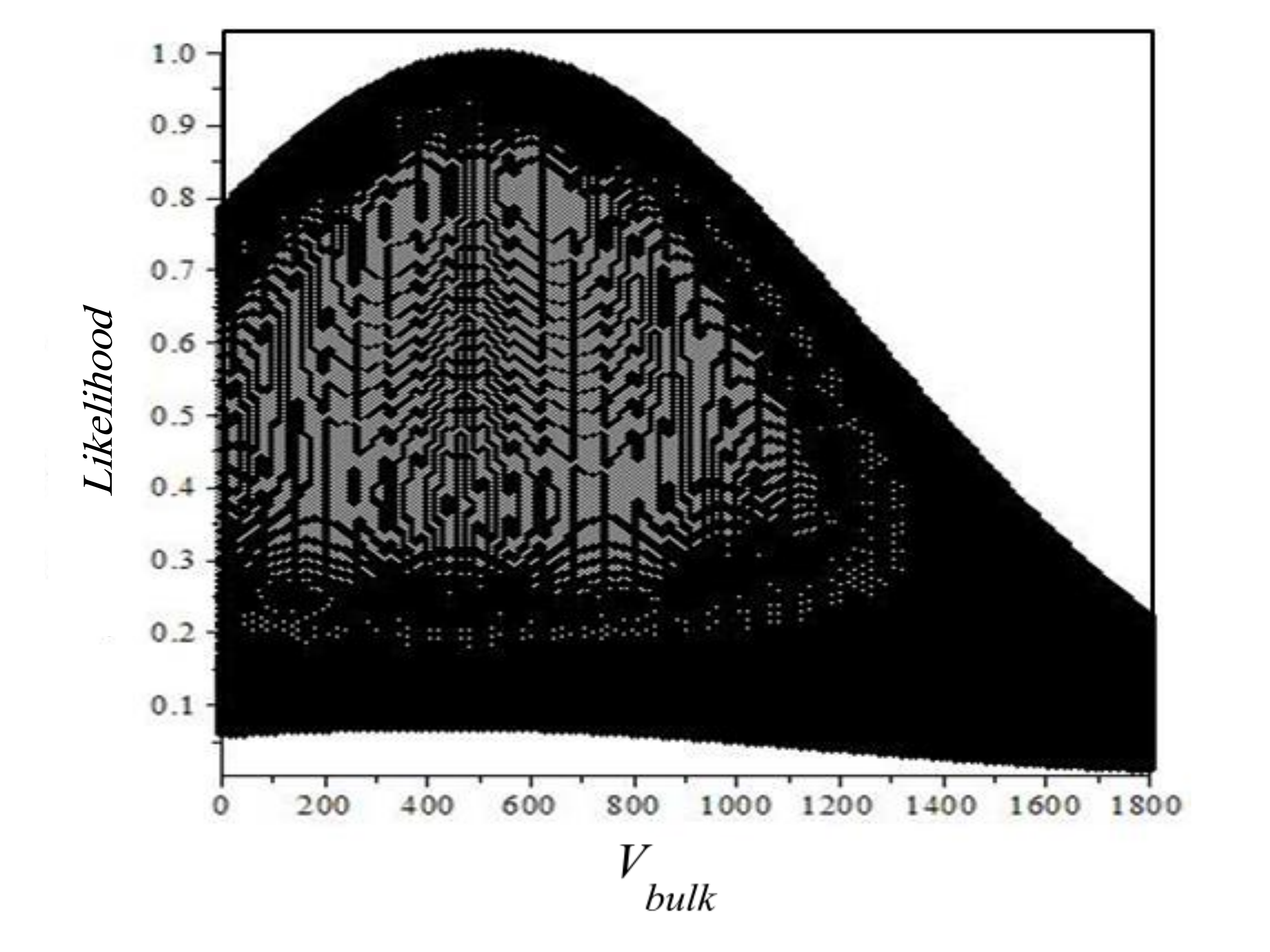}{0.5\textwidth}{(a)}
          \fig{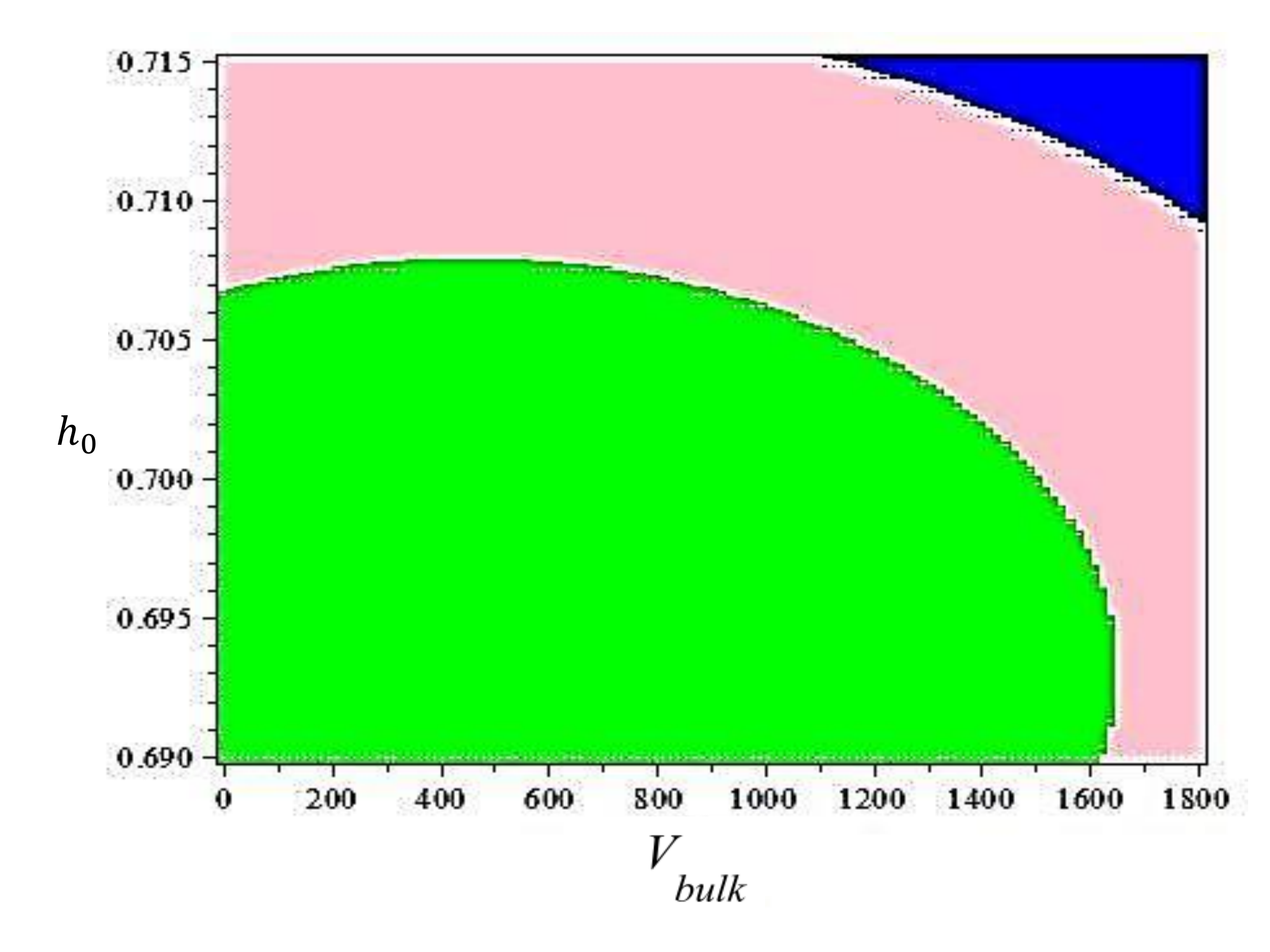}{0.5\textwidth}{(b)}
          }
\gridline{\fig{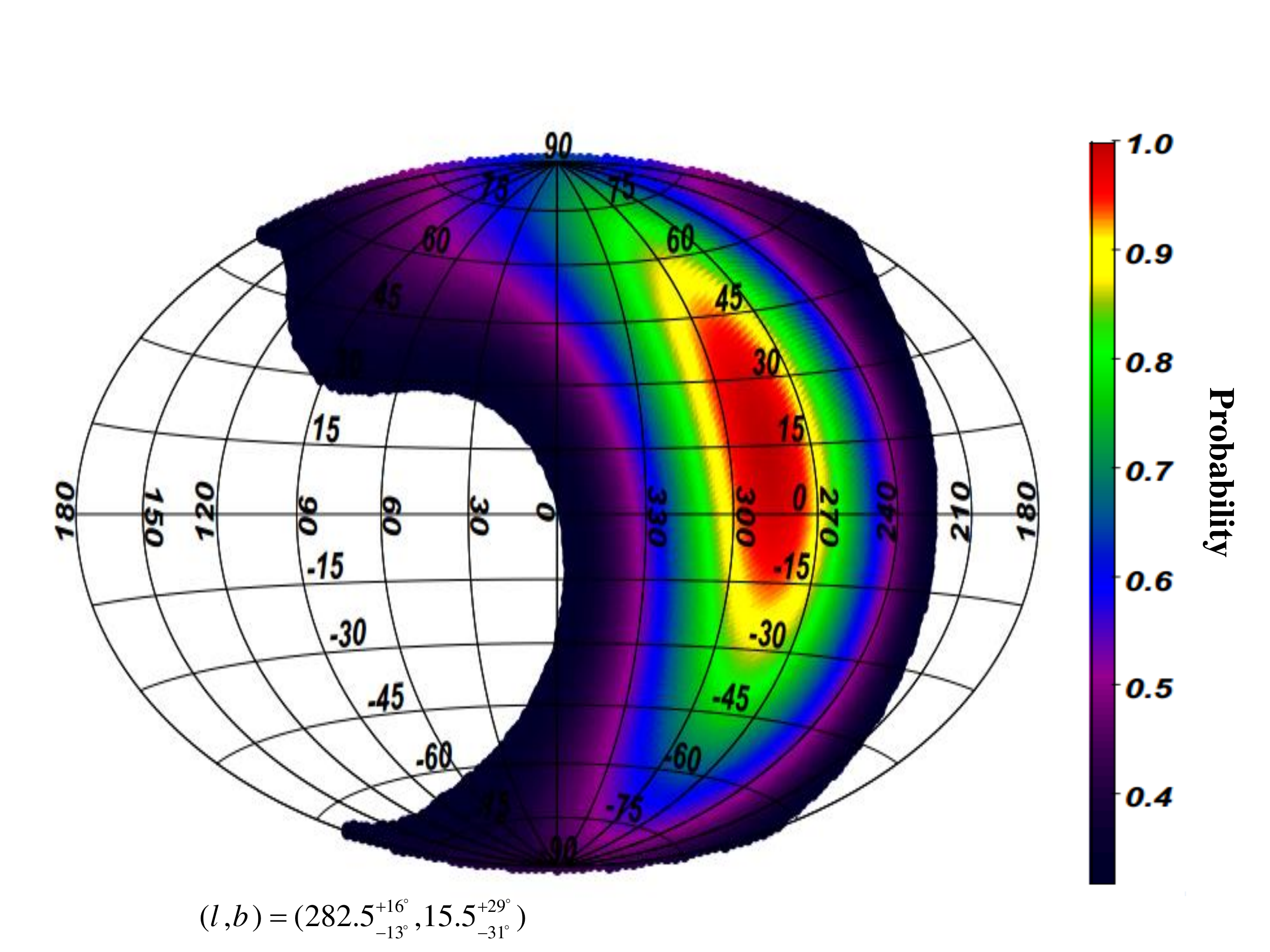}{0.97\textwidth}{(c)}
          }
\caption{\small{\emph{Probability of bulk flow direction in galactic longitude $l$ and galactic
latitude $b$ for $0.06<z<0.1$.The most probable direction pointing towards$(l; b) = (282.5^{+16^{o}}_{-13^{o}}; 15.5^{+29^{o}}_{-31^{o}})$. Distribution of SNe Ia on the sky in galactic coordinates. The results of other studies have also been shown }}.\label{fig:p8}}
\end{figure*}

\begin{figure*}
\gridline{\fig{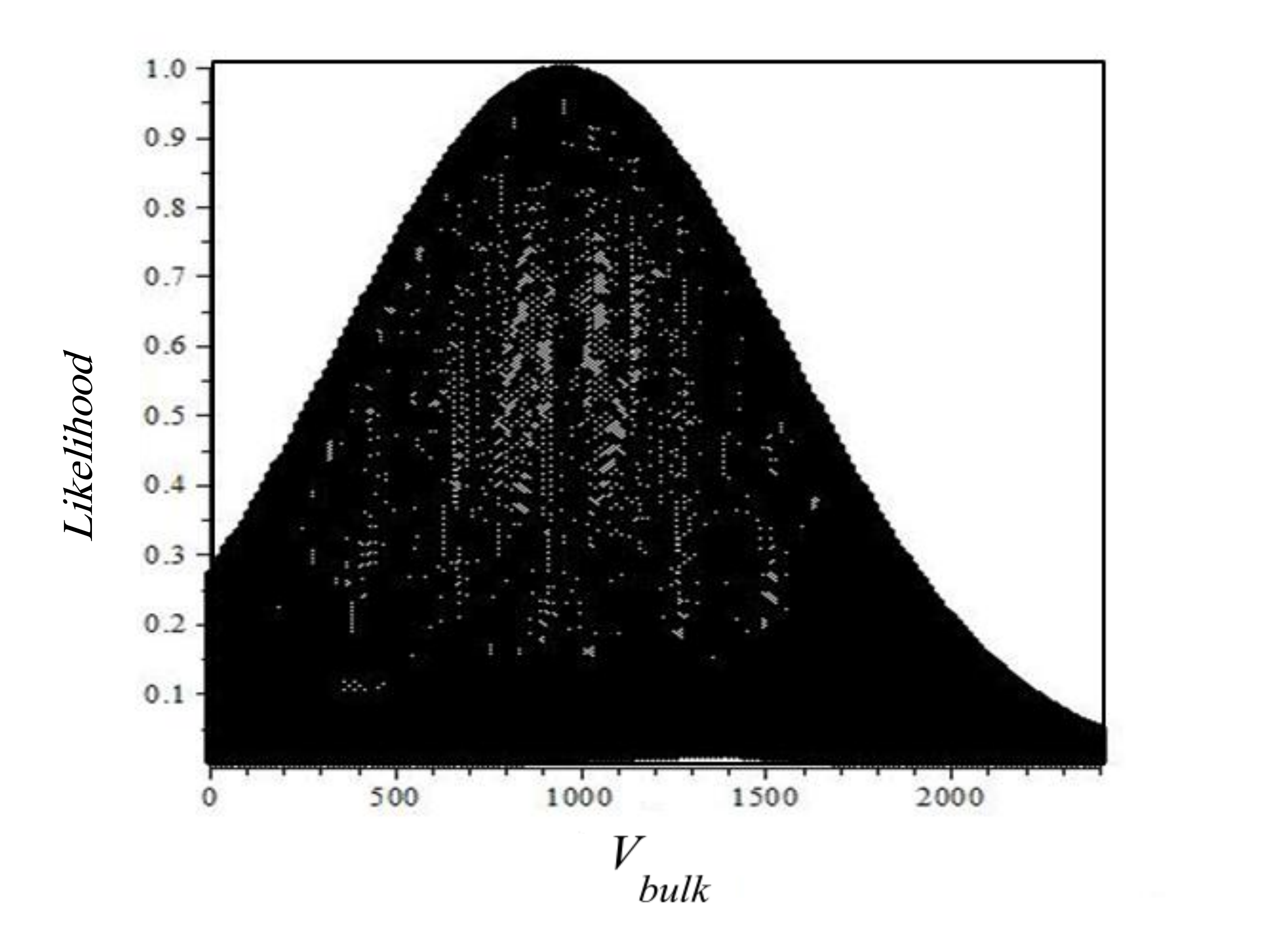}{0.5\textwidth}{(a)}
          \fig{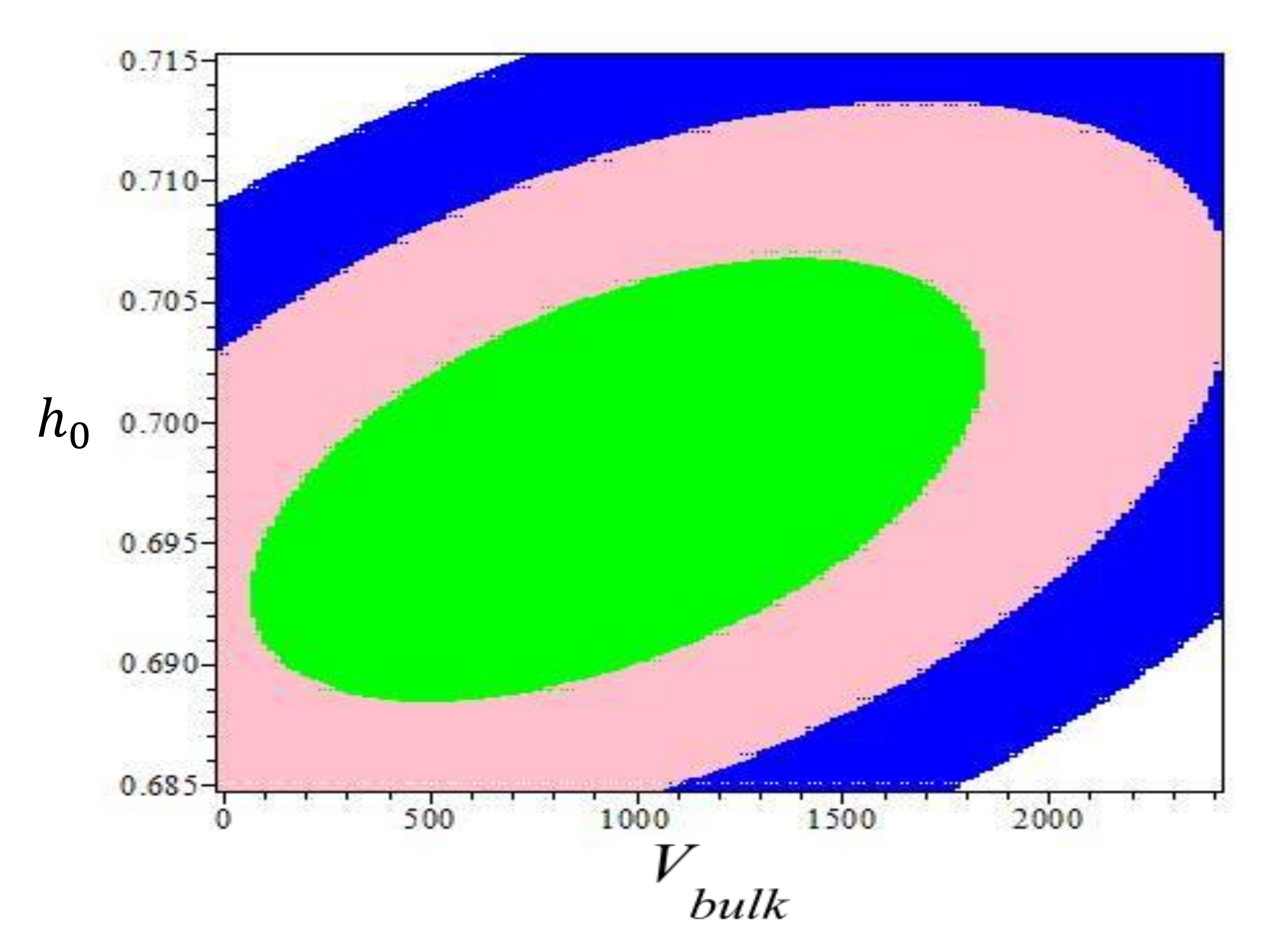}{0.5\textwidth}{(b)}
          }
\gridline{\fig{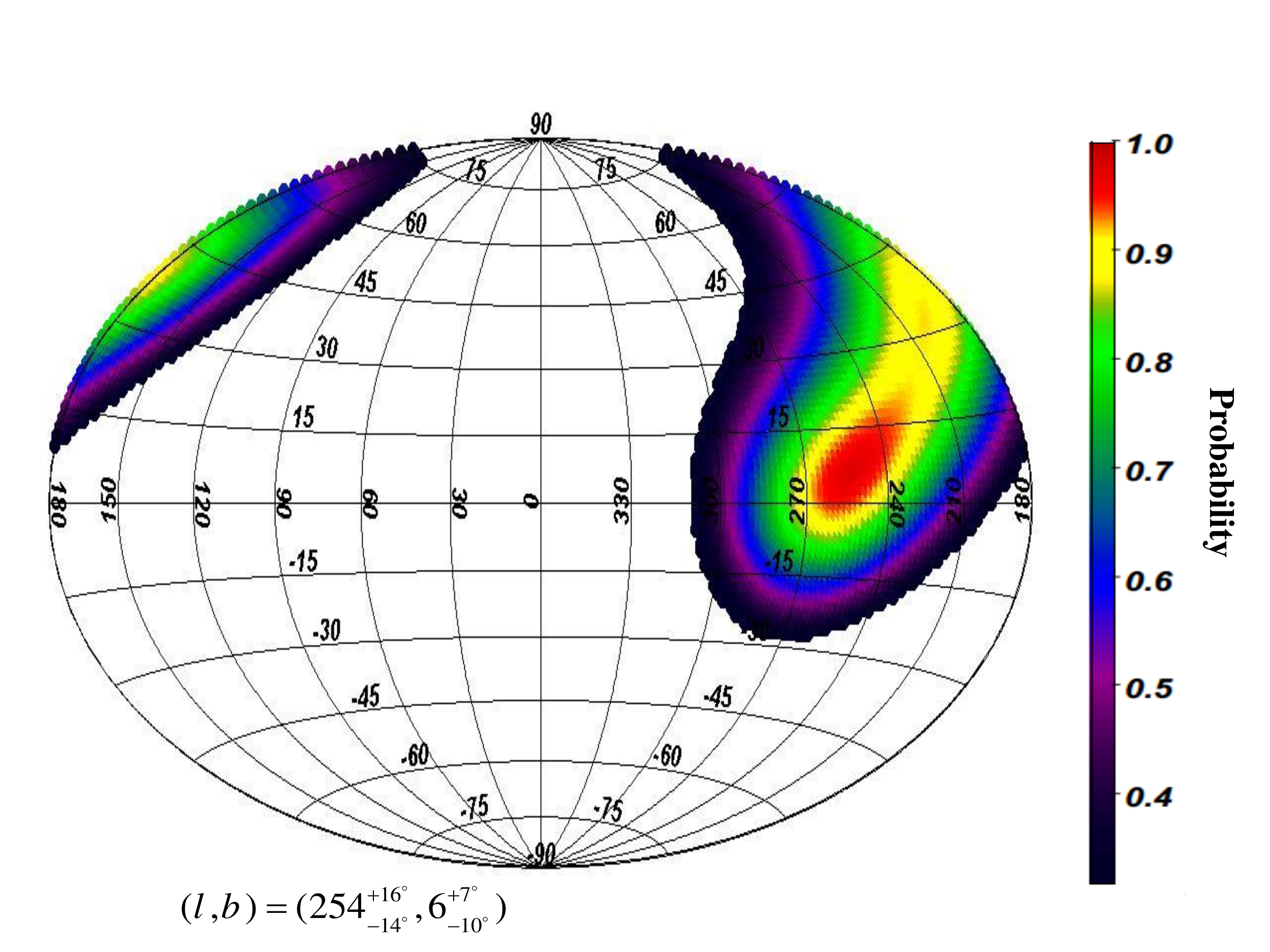}{0.97\textwidth}{(c)}
          }
\caption{\small{\emph{Probability of bulk flow direction in galactic longitude $l$ and galactic
latitude $b$ for $0.1<z<0.2$.The most probable direction pointing towards$(l; b) = (254^{+16^{o}}_{-14^{o}}; 6^{+7^{o}}_{-10^{o}})$. Distribution of SNe Ia on the sky in galactic coordinates. The results of other studies have also been shown }}.\label{fig:p9}}
\end{figure*}

\subsection{Redshift tomography}
As we mentioned in previous section, since the high-redshift results may be contaminated by the lowred shift data, we perform a ‘cosmic tomography’ in which the data are sliced up in redshift and the question of isotropy is studied separately
for each slice. Our results for $QCDM$ model are
summarized in Table I and  for three important redshift range $0.035<z<0.0, 0.06<z<0.1$ and $0.1<z<0.2$, they have been depicted in Figs(\ref{fig:p7}) to (\ref{fig:p9})   There are interesting results in redshift tomography.\\\\
$\bullet$ The results of direction and amplitude of bulk flow have been obtained for a slice are much different from those obtained for cumulative redshift slices of the data.\\
$\bullet$ Surprisingly, for high redshift sells $z>0.035$, we found a larger amplitude flow; $v_{bulk}\simeq 500-1000 kms^{-1}$ which is in excellent agrement with
the results of \citep{Kashlinsky1}-\citep{Kashlinsky2}-\citep{Kashlinsky3}-\citep{Kashlinsky4}
nearly. \\
$\bullet$ Recently \citep{Colin} investigated
anisotropies in discrete redshift shells using the Union2 compilation of Type Ia SNe \citep{Amanullah}(The data have been used in this paper). Although our results are in excellent agreement with in \citep{Colin} low redshift $z<0.1$ , however in high redshifts $z>0.1$, our results are different. In contrast to \citep{Colin} who found that in high redshifts the agreement between the SNe Ia data and the $\Lambda CDM$ model does improve, we found that contradiction between $\Lambda CDM$ and SNe Ia data is revealed more in high redshifts $z>0.1$. Because of using same data, we can conclude that the disagreement between the results reefers to different background cosmological models($\Lambda CDM, QCDM$) which have a degenerate behavior in low redshifts and it would be break at high redshifts. In other word, "\emph{quintessence behaves as a smooth component: it does not participate directly in cluster formation, but it only alters the background cosmic evolution, however  at very large scales$ (\sim > 100 h^{-1}Mpc)$, Quintessence clusters gravitationally, leaving an imprint on the microwave background anisotropy}"\citep{Caldwell}.

\begin{table}
\renewcommand{\thetable}{\arabic{table}}
\centering
\caption{The comparison of achieved Bulk velocity in this paper with other studies for dark energy dipole} \label{tab:decimal}
\begin{tabular}{c@{\hspace{6mm}}c@{\hspace{6mm}}c@{\hspace{6mm}}c@{\hspace{6mm}}c@{\hspace{6mm}}c@{\hspace{6mm}}c@{\hspace{6mm}} c@{\hspace{6mm}}c@{\hspace{6mm}}c@{\hspace{6mm}}c@{\hspace{6mm}}} % centered columns (5 columns)
\hline\hline %inserts double horizontal lines
Range &\ \ $l^{o}$&\ \ $b^{o}$ &$ x_{0}$ & $y_{0}$& $\lambda$& $h_{0}$& $v_{bulk}$ & $\chi^{2}_{min}$& data\\
\rule{0pt}{8mm}%
 &\ \ &\ \  & & & &$km s^{-1}Mpc^{-1}$ & $km s^{-1}$ & & Number\\
[3ex] % inserts table
%heading
\hline % inserts single horizontal line
$0.015<z<0.025$ &$244.5^{+19}_{-19}$ &$-19^{+15}_{-15}$ & $0.425$ & $0.834$ & $2.32$ & $0.698$ & $300$ & $46.1724321$ &  58 \\
\rule{0pt}{8mm}%
$0.015<z<0.035$  &$292^{+22}_{-22}$ &$10.5^{+17}_{-17}$ & $0.425$ & $0.996$ & $-0.098$ & $0.697$ & $268$ & $97.97402438$ &109 \\
\rule{0pt}{8mm}%
$0.015<z<0.06$ &$300^{+18}_{-18}$ &$6^{+14}_{-14}$ & $0.451$ & $0.896$ & $-0.078$ & $0.699$ & $257$ & $130.4102306$ & 142  \\
\rule{0pt}{8mm}%
$0.015<z<0.1$  &$ 302^{+20}_{-20}$ &$3^{+10}_{-10}$ & $0.560$ & $0.981$ & $-0.095$ & $0.698$ &  $246$ & $148.5758203$ & 165   \\
\rule{0pt}{8mm}%
$0.015<z<1.4$  &$ 296^{+34.6}_{-34.6}$ &$1^{+23.5}_{-23.5}$ & $0.395$ & $0.839$ & $2.2$ & $0.697$ &  $253$ & $530.61029$ & 556   \\
\rule{0pt}{8mm}%
$0.035<z<0.06$  &$ 4.8^{+12}_{-18}$ &$-14.5^{+12}_{-10}$ & $0.161$ & $0.796$ & $2.1$ & $0.702$ &  $858$ & $27.36652794$ & 33 \\
\rule{0pt}{8mm}%
$0.06<z<0.1$ &$ 282.5^{+16}_{-13}$ &$15.5^{+29}_{-31}$ & $0.773$ & $0.996$ & $2$ & $0.701$ &  $519$ & $18.33498129$ & 23 \\
\rule{0pt}{8mm}%
$0.1<z<0.2$ &$ 254^{+16}_{-14}$ &$6^{+7}_{-10}$ & $0.2$ & $0.799$ & $1$ & $0.6976$ &  $1014$ & $53.256270$ & 55 \\
\rule{0pt}{8mm}%
$0.2<z<0.4$ &$ 138^{+17}_{-14}$ &$24^{+14}_{-11}$ & $0.259$ & $0.848$ & $1.5$ & $0.6972$ &  $1200$ & $108.5668242$ & 124\\
\rule{0pt}{8mm}%
$0.4<z<0.6$ &$ 300^{+19}_{-15}$ &$60^{+17}_{-14}$ & $0.27$ & $0.859$ & $2.5$ & $0.6974$ &  $597$ & $100.728441$ & 101 \\
\rule{0pt}{8mm}%
$0.6<z<0.8$ &$ 202.5^{+18}_{-14}$ &$81^{+17}_{-12}$ & $0.26$ & $0.837$ & $1.4$ & $0.7014$ &  $1200$ & $47.07531$ & 50  \\
\rule{0pt}{8mm}%
$0.8<z<1$ &$ 360^{+15}_{-11}$ &$30^{+14}_{-10}$ & $0.349$ & $0.829$ & $2.2$ & $0.701$ &  $570$ & $47.59088633$ & 40  \\
\rule{0pt}{8mm}%
$0.1<z<1$ &$ 280.5^{+12}_{-13}$ &$-15^{+8}_{-7}$ & $0.346$ & $0.834$ & $2.29$ & $0.699$ &  $1050$ & $364.5724654$ & 372  \\
 % inserting body of the table
 [1ex] % [1ex] adds vertical space
\hline %inserts double horizontal lines
\end{tabular}
\end{table}

\section{Conclusion}
Previous studies of bulk flow can be classified in two set of results. Some studies reported possible large bulk flows at scales of $\sim 100 h^{-1}Mpc$, \citep{Hudson}; \citep{Kashlinsky}; \citep{Watkins}; \citep{Lavaux}; \citep{Dai}; \citep{Colin}; \citep{Macaulay}; \citep{Feindt} while others reported bulk flow to be consistent with the expectation from $\Lambda CDM $; \citep{Courteau}; \citep{Nusser}; \citep{Nusser1}; \citep{Branchini}; \citep{Turnbull}; \citep{Ma}]. Over larger distances, \citep{Kashlinsky} reported a bulk flow out to $d\geq 300 h^{-1}Mpc$ . According to their results \citep{Kashlinsky}-\citep{Kashlinsky4} the bulk flow is $\sim1000 km s^{-1}$ in the direction of the CMB dipole up to a distance of at least $\sim800 h^{-1}Mpc$. A flow of this amplitude on such a large scale is in contradict with that predicted by the $\Lambda CDM$ (one-dimensional rms
velocity is $\sim110 kms^{-1}$)\citep{Watkins}.
In this paper, we study the bulk flow of the local universe using Type Ia supernova data in QCDM model. We find that at low redshift bulk flow is moving towards the $(l; b) = (302^{o} \pm20^{o}; 3^{o}\pm10^{o})$ direction with $v _{bulk} = 240\pm 25kms^{-1} $ velocity. This direction is aligned  with direction of (SSC) and agreement with a number previous studies at $(1-\sigma)$, however for high redshift we get $v _{bulk} = 1000\pm 25kms^{-1} $ towards the $(l; b) = (302^{o} \pm20^{o}; 3^{o}\pm10^{o})$. This indicates that for low redshift our results are approximately consistent with the $\Lambda CDM$ model with the latest WMAP best fit cosmological parameters, however for high redshift they are in disagreement of $\Lambda CDM$ and support the results of previous studies such as kashlinsky et al which report the large bulk flow for the Universe.\\
There are several possible explanations for the discrepancy we
have observed;\\\\
 $\bullet$ Our results are in excellent agreement with in \citep{Colin} low redshift $z<0.1$ , however in high redshifts $z>0.1$, our results are different. In contrast to \citep{Colin} who found that in high redshifts the agreement between the SNe Ia data and the $\Lambda CDM$ model does improve, we found that contradiction between $\Lambda CDM$ and SNe Ia data is revealed more in high redshifts $z>0.1$. Because of using same data, we can conclude that the disagreement between the results reefers to different background cosmological models($\Lambda CDM, QCDM$) which have a degenerate behavior in low redshifts and it would be break at high redshifts. In other word, "\emph{quintessence behaves as a smooth component: it does not participate directly in cluster formation, but it only alters the background cosmic evolution, however  at very large scales$ (\sim > 100 h^{-1}Mpc)$ and leaving an imprint on the microwave background anisotropy}" \citep{Caldwell}.\\\\
$ \bullet$ We can conclude that at small scales, fluctuations in the dark energy are damped and do not enter
in the evolution equation for the perturbations in the pressureless matter. Thus quintessence behaves as a smooth component: it does not participate directly in cluster formation, but it only alters the background cosmic evolution, however  at very large scales$ (\sim > 100 h^{-1}Mpc)$, Quintessence clusters gravitationally, leaving an imprint on the microwave background anisotropy. In other world, quintessence remains smooth like the cosmological constant on small length scales. The quintessence fluctuations are weak compared with the matter
fluctuations at smaller scales.\\\\
$ \bullet$  While the direction of the flow from different works agrees well, there is considerable variation in the magnitude of the
flow. Part of the discrepancy between the results may be related to this fact that magnitude of the flow can depend strongly on the depth of the survey. For example comparison of results of \citep{Kashlinsky} and \citep{Watkins} shows that the direction of bulk flow in two studies are in excellent agrement , while the amplitude of their flow are considerately different. Note that \citep{Kashlinsky} sample(volume of radius of $\sim120-600 h^{-1}Mpc$)
is very much deeper than \citep{Watkins} (volume of radius of $\sim100 h^{-1}Mpc$)\\\\
$ \bullet$ We found for each slice of data which contain low redshift (even the large slice with $0.015<z<1.4$), the amplitude of  $v_{bulk}$ is close to $250 kms^{-1}$.Thus we performed a ‘cosmic tomography’  where the data are sliced up in redshift and the question of isotropy was studied separately. Surprisingly, for high redshift sells $z>0.35$, we found a larger amplitude flow; $v_{bulk}\simeq 500-1000 kms^{-1}$ which is in excellent agrement with
the results of \citep{Kashlinsky1}-\citep{Kashlinsky2}-\citep{Kashlinsky3}-\citep{Kashlinsky4}
nearly. This indicates that, due to sparseness of the data at high redshift, the high-redshift results may be contaminated by the low redshift data. Also
at low redshifts $z<< 1$, the Hubble law indicates a
linear relationship between distance and redshift so the choice of cosmological model is irrelevant; however this becomes important at high redshift \citep{Colin}.\\\\
$ \bullet$  It is possible that the large
observed flow is the result of a systematic error in the data, although
the independence of the distance indicators (TF, FP and SN Ia) and
methodology of the various surveys, as well as the agreement between different surveys makes this unlikely \citep{Watkins}\\
$ \bullet$ Cluster evolution offers a promising approach for breaking the degeneracy\\
$ \bullet$ While the quintessence fluctuations are weak compared with the matter
fluctuations at smaller scales and the quintessence energy density is negligible when
those length-scale enter the horizon, however, these fluctuations have a non negligible effect on the cosmic microwave background
anisotropy and the mass power spectrum \cite{paul}\\\\\\




\begin{thebibliography}{}
\expandafter\ifx\csname natexlab\endcsname\relax\def\natexlab#1{#1}\fi
\providecommand{\url}[1]{\href{#1}{#1}}
\providecommand{\dodoi}[1]{doi:~\href{http://doi.org/#1}{\nolinkurl{#1}}}
\providecommand{\doeprint}[1]{\href{http://ascl.net/#1}{\nolinkurl{http://ascl%
.net/#1}}}
\providecommand{\doarXiv}[1]{\href{https://arxiv.org/abs/#1}{\nolinkurl{https:%
//arxiv.org/abs/#1}}}

\end{thebibliography}


\begin{thebibliography}{}
\bibitem [Lineweaver(1996)]{Lineweaver} C. H. Lineweaver, in Microwave background anisotropies.
Proceedings, 31st Rencontres de Moriond, 16th Moriond
Astrophysics Meeting, Les Arcs, France, March 16-23,
1996 (1997) pp. 69–76, arXiv:astro-ph/9609034 [astroph].
\bibitem [Conklin(1969)]{Conklin} E.K.Conklin, Nature, 222, 971 (1969)
\bibitem [Henry(1971)]{Henry} Henry P. S., 1971, Nature, 231, 516
\bibitem [Kogut et al (1993)] {Kogut} A. Kogut  et al., 1993, ApJ, 419, 1
\bibitem [Yahil et al (1980)] {Yahil} Yahil, A., Sandage, A., \& Tamman, G. A. 1980, ApJ, 242, 448
\bibitem [Davis et al (1982)]{Davis0} Davis, M., \& Huchra, J. 1982, ApJ, 254, 437
\bibitem [Yahil et al (1986)]{Yahil2} Yahil, A., Walker, D., \& Rowan-Robinson, M. 1986, ApJ, 301, L1
\bibitem [Lynden-Bell et al (1989)]{Lynden} Lynden-Bell, D., Lahav, O., \& Burstein, D. 1989, MNRAS, 241, 325
\bibitem [Davis \& Peebles(1983)] {Davis} Davis, M., \& Peebles, J. 1983, ARA\&A, 21,109
\bibitem [Villumsen \& Strauss(1987)]{Villumsen} Villumsen, J., \& Strauss, M. 1987, ApJ, 322, 37
\bibitem [Aaronson et al(1986)] {Aaronson} Aaronson, M., Bothun, G., Mould, J., Huchra, J., Schommer, R., \& Cornell, M.1986, ApJ, 302, 536
\bibitem [Shaya (1984)] {Shaya} Shaya,E.1984,ApJ280,470
\bibitem [Tammann \& Sandage(1985)] {Tammann} Tammann,G.and Sandage,A.1985,ApJ294,81
\bibitem [Lijle et al(1987)] {Lijle} Lijle, P., Yahil, A., \& Jones, B. T. 1987, ApJ, 307,91
\bibitem [Aaronson et al(1982)] {Aaronson1} Aaronson, M., Huchra, J., Mould, J., Schechter, P. L., \& Tully, R. B. 1982, ApJ,258,64
\bibitem [Han \& Mould(1990)] {Han} Han, H., \& Mould, J. 1990, ApJ, 360,448
\bibitem [Lahav et al(1990)] {Lahav} Lahav, O., Kaiser, N., \& Hoffman, Y. 1990, ApJ, 352,448
\bibitem [Dressier et al(1987)] {Dressier} Dressier, A., Faber, S., Burstein, D., Davies, R., Lynden-Bell, D., Terlevich, R.,
\& Wegner, G. 1987, ApJ, 313, L37
\bibitem [Lynden-Bell et al(1988)] {Lynden-Bell} Lynden-Bell et al.1988,ApJ326,19
\bibitem [Robinson et al (1990)]{Robinson}Rowan-Robinson, M., et al. 1990, MNRAS, 247, 1
\bibitem [Strauss et al (1992)] {Strauss}Strauss, M. A., et al. 1992, ApJ, 83, 29
\bibitem [Hudson et al (1993)] {Hudson} Hudson, M. J. 1993, MNRAS, 265, 72
\bibitem [Courteau et al(2000)] {Courteau} Courteau, S., Willick, J. A., Strauss, M. A., Schlegel, D., \& Postman, M. 2000,
ApJ, 544, 636
\bibitem [Caldwell et al (1998)] {Caldwell} Caldwell R.R., Dave R., Steinhardt P.J., 1998, Phys. Rev. Lett. 80, 1582
\bibitem [Nusser \& Davis(2011)] {Nusser} Nusser, A. \& Davis, M. 2011, ApJ, 736, 93
\bibitem [Nusser et al(2011)] {Nusser1} Nusser, A., Branchini, E., \& Davis, M. 2011, ApJ, 735, 77
\bibitem [Branchini et al(2012)] {Branchini} Branchini, E., Davis, M., \& Nusser, A. 2012, MNRAS, 424, 472
\bibitem [Turnbull et al(2012)] {Turnbull}Turnbull, S. J., Hudson, M. J., Feldman, H. A., et al. 2012, MNRAS, 420, 447
\bibitem [Ma \& Scott(2013)] {Ma} Ma, Y.-Z. \& Scott, D. 2013, MNRAS, 428, 2017
\bibitem [Hudson et al(2004)] {Hudson} Hudson, M. J., Smith, R. J., Lucey, J. R., \& Branchini, E. 2004, MNRAS, 352,61
\bibitem [Kashlinsky et al (2008)] {Kashlinsky} A. Kashlinsky; F. Atrio-Barandela, D. Kocevski, and H. Ebeling, Astrophys. J. 686, L49
(2008).
\bibitem [Watkin et al(2009)] {Watkins} Watkins, R., Feldman, H. A., \& Hudson, M. J. 2009, MNRAS, 392, 743
\bibitem [Lavaux et al(2010)] {Lavaux} Lavaux, G., Tully, R. B., Mohayaee, R., \& Colombi, S. 2010, ApJ, 709, 483
\bibitem [Lavaux et al(2013)] {Lavaux2}MNRAS 430, 1617–1635 (2013)
\bibitem [Dai et al(2011)] {Dai} Dai, D.-C., Kinney, W. H., \& Stojkovic, D. 2011, J. Cosmol. Astropart. Phys., JCAP04(2011)015
\bibitem [Colin et al(2011)] {Colin} Colin, J., Mohayaee, R., Sarkar, S., \& Shafieloo, A. 2011, MNRAS, 414, 264
\bibitem [Feindt et al(2013)] Feindt et al, A and A 560, A90 (2013)
\bibitem [Macaulay et al(2012)] {Macaulay} Macaulay, E., Feldman, H. A., Ferreira, P. G., et al. 2012, MNRAS, 425, 1709
\bibitem [Feindt et al(2013)]  {Feindt} U. Feindt et al. A \& A, 560, A90 (2013). DOI:10.1051/0004-
6361/201321880, arXiv:1310.4184v3 [astro-ph.CO] 1 Jul 2015
\bibitem [Kashlinsky et al (2009)] {Kashlinsky1} Kashlinsky, A., Atrio-Barandela, F., Kocevski, D., \& Ebeling, H. 2009, ApJ, 691, 1479
\bibitem [Kashlinsky et al(2010)] {Kashlinsky2} A. Kashlinsky; F. Atrio-Barandela, H. Ebeling, A. Edge, and D. Kocevski, Astrophys. J. 712,
L81 (2010).
\bibitem [Kashlinsky et al(2011)] {Kashlinsky3} Kashlinsky, A., Atrio-Barandela, F., \& Ebeling, H. 2011, ApJ, 732, 1
\bibitem [Kashlinsky et al(2012)] {Kashlinsky4} Kashlinsky, A., Atrio-Barandela, F., \& Ebeling, H. 2012, ArXiv e-prints
\bibitem [Amanullah et al(2010)] {Amanullah} Amanullah, R., Lidman, C., Rubin, D., et al. 2010, ApJ, 716, 712
\bibitem [Ratra \& Peebles et al(1988)] {Ratra} Ratra B., Peebles P.J.E., 1988, Phys.Rev. D 37, 3406
\bibitem [Sunyaev \& Zeldovich (1972)] {Sunyaev} R.A. Sunyaev and Ya.B. Zeldovich, Comments on Astrophysics and Space Physics 4, 173
(1972).
\bibitem [Sunyaev \& Zeldovich et al(1980)] {Zeldovich} R.A. Sunyaev and Ya.B. Zeldovich, MNRAS 190, 413 (1980).
\bibitem [Kocevski \& Ebeling(2006)] {Kocevski} Kocevski, D. D. \& Ebeling, H. 2006, ApJ, 645, 1043
\bibitem [Shapley (1930)] {Shapley} Shapley, H. 1930, Harvard College Observatory Bulletin, 874, 9
\bibitem [Scaramella et al(1989)] {Scaramella} Scaramella, R., Baiesi-Pillastrini, G., Chincarini, G., Vettolani, G., \& Zamorani,
G. 1989, Nature, 338, 562
\bibitem [Raychaudhury et al(1991)] {Raychaudhury} Raychaudhury, S., Fabian, A. C., Edge, A. C., Jones, C., \& Forman, W. 1991,
MNRAS, 248, 101
\bibitem [Schwarz \& Weinhorst (2007)] {Schwarz} Schwarz, D. J. \& Weinhorst, B. 2007, A\&A, 474, 717
\bibitem [Kalus et al(2013)] {Kalus} Kalus, B., Schwarz, D. J., Seikel, M., \& Wiegand, A. 2013, accepted, A\&A
\bibitem [Sasaki (1987)] {Sasaki} Sasaki Mon.Not. Roy. Astron. Soc. 228, 653 (1987)
\bibitem [Pyne \&  Birikinshaw (2004)] {Pyne} T. Pyne ,  M. Birikinshaw,  Mon.Not.Roy.Astron.Soc. 384 (2004) 581
\bibitem [Hui \& Greene(2006)] {Hui} Hui, L. \& Greene, P. B. 2006, Phys.Rev., D73, 123526
\bibitem [Bolejko et al(2013)] {Bolejko} Krzysztof Bolejko, Chris Ciarkson, Roy Maartens, David Bacon, Nikolai Meures, Emma Beynon, Phys. Rev. LETT. 110.021302(2013)
 \bibitem [Weinberg(1989)] {Weinberg} S. Weinberg, Rev. Mod. Phys. 61 (1989) 1.
\bibitem [Peebles et al(2003)] {Peebles} P.J.E. Peebles, B. Ratra, Rev. Mod. Phys. 75 (2003) 559.
\bibitem [Bonvin et al(2006)] {Bonvin} Bonvin, C., Durrer, R., \& Kunz, M. 2006, Physical Review Letters, 96, 191302
\bibitem [Feldman et al(2010)] {Feldman} Feldman H. A., Watkins R., Hudson M. J., 2010, MNRAS, 407,
2328
\bibitem [Wang \& Wang (2014)]  {Wang} J.S. Wang and F.Y. Wang, Probing the anisotropic expansion from supernovae and GRBs in
a model-independent way, Mon. Not. Roy. Astron. Soc. 443 (2014) 1680 [arXiv:1406.6448][INSPIRE].
\bibitem [Mariano \& Perivolaropoulos (2012)] {Mariano} A. Mariano and L. Perivolaropoulos, Is there correlation between Fine Structure and Dark
Energy Cosmic Dipoles, Phys. Rev. D 86 (2012) 083517 [arXiv:1206.4055] [INSPIRE].
\bibitem [Chang et al(2013)] {Chang} Z. Chang, M.-H. Li, X. Li and S. Wang, Cosmological model with local symmetry of very
special relativity and constraints on it from supernovae, Eur. Phys. J. C 73 (2013) 2459[arXiv:1303.1593] [INSPIRE].
\bibitem [Yang et al(2014)]  {Yang}  X. Yang, F.Y. Wang and Z. Chu, Searching for a preferred direction with Union2.1 data,
Mon. Not. Roy. Astron. Soc. 437 (2014) 1840 [arXiv:1310.5211] [INSPIRE].
\bibitem [Cai et al (2013)] {Cai} Rong-Gen Cai, Yin-Zhe Ma, Bo Tang, Zhong-Liang Tuo, Phys. Rev. D 87, 123522 (2013) arXiv:1303.0961v4 [astro-ph.CO]
\bibitem [Salehi \& Aftabi (2016)] {Salehi} A. Salehi, S. Aftabi, JHEP 1609(2016)140 arXiv:1502.04507v4 [gr-qc]
\bibitem [Ma et al(2011)] {Ma1} Ma Y.-Z., Gordon C., Feldman H. A., 2011, Phys. Rev. D., 83, 103002
\bibitem [Hoffman  et al(2001)] {Hoffman} Hoffman Y., Eldar A., Zaroubi S., Dekel A., 2001, preprint (astro-ph/
0102190)
\bibitem [Hudson (1994b)] {Hudson1} Hudson M. J., 1994b, MNRAS, 266, 475
\bibitem [Pike \& Hudson (2005)] {Pike} Pike R. W., Hudson M. J., 2005, ApJ, 635, 11

\bibitem [Erdogdu et al(2006)] {Erdogdu} P. Erdogdu, J.P. Huchra, O. Lahav, M. Colless, R.M. Cutri, E. Falco, T. George, T. Jarrett, D. H. Jones, C.S. Kochanek, L. Macri, J. Mader, N. Martimbeau, M. Pahre, Q. Parker, A. Rassat, W. Saunders, 2006 arXiv:astro- ph/0507166
\bibitem [Huchra et al(2011)] {Huchra} John P. Huchra, Lucas M. Macri, Karen L. Masters, Thomas H. Jarrett, Perry Berlind, Michael Calkins, Aidan C. Crook, Roc Cutri, Pirin Erdogdu, Emilio Falco, Teddy George, Conrad M. Hutcheson, Ofer Lahav, Jeff Mader, Jessica D. Mink, Nathalie Martimbeau, Stephen Schneider, Michael Skrutskie, Susan Tokarz, Michael Westover, 2011arXiv:1108.0669[astro-ph.CO]
\bibitem [Giovanelli et al(1996)] {Giovanelli} R. Giovanelli, M. Haynes , T. Herter , N. Vogt , L. da Costa , W. Freudling , J. Salzer , G. Wegner, 1996 arXiv:astro-ph/9610117
\bibitem [Springob et al(2007)] {Springob} Christopher M. Springob, Karen L. Masters, Martha P. Haynes, Riccardo Giovanelli, Christian Marinoni,2007 APJS, 172: 599-614.





\bibitem [Smoot et al(1977)] {Smoot} Smoot, G. F., Gorenstein, M. V., \& Muller, R. A. 1977, Phys. Rev. Lett., 39,898
\bibitem [Cheng et al(1979)] {Cheng1} Cheng, E. S., Saulson, P. R., Wilkinson, D. T, \& Corey, B. E. 1979, ApJ, 232,L139
\bibitem [Amanullah et al(2010)] {Amanullah} Amanullah, R., Lidman, C., Rubin, D., et al. 2010, ApJ, 716, 712
\bibitem [Steinhardt (2003)]{paul} Steinhardt PJ, Philos Trans A Math Phys Eng Sci. 2003 Nov 15;361(1812):2497-513.



\end{thebibliography}
\end{document}